# 2024 Roadmap on 2D Topological Insulators

Bent Weber[1,*], Michael S. Fuhrer[2,3,*], Xian-Lei Sheng[4], Shengyuan A. Yang[5], Ronny Thomale[6], Saquib Shamim[7,8,9], Laurens W. Molenkamp[7,8], David Cobden[10], Dmytro Pesin[11], Harold J. W. Zandvliet[12], Pantelis Bampoulis[12], Ralph Claessen[13], Fabian R. Menges[14], Johannes Gooth[14], Claudia Felser[14,25], Chandra Shekhar[14], Anton Tadich[15], Mengting Zhao[16], Mark T. Edmonds[16], Junxiang Jia[1], Maciej Bieniek[17,18], Jukka I. Väyrynen[19], Dimitrie Culcer[20,21], Bhaskaran Muralidharan[22], Muhammad Nadeem[23,24]

1. Nanyang Technological University
2. School of Physics and Astronomy, Monash University, Clayton, Victoria 3800, Australia
3. ARC Centre of Excellence in Future Low-Energy Electronics Technologies (FLEET), Monash University, Clayton, Victoria 3800, Australia
4. School of Physics, Beihang University, China
5. Research Laboratory for Quantum Materials, Singapore University of Technology and Design, Singapore
6. Institut für Theoretische Physik und Astrophysik and Würzburg-Dresden Cluster of Excellence ct.qmat, Julius-Maximilians-Universität Würzburg, Germany
7. Experimentelle Physik III, Physikalisches Institut, Universität Würzburg, Am Hubland, 97074, Würzburg, Germany
8. Institute for Topological Insulators, Universität Würzburg, Am Hubland, 97074, Würzburg, Germany
9. Department of Condensed Matter and Material Physics, S. N. Bose National Centre for Basic Sciences, Kolkata 700106, India
10. University of Washington, Seattle
11. Department of Physics, University of Virginia, Charlottesville, VA 22904-4714
12. Physics of Interfaces and Nanomaterials, MESA+ Institute for Nanotechnology, University of Twente, P.O. Box 217, 7500AE Enschede, The Netherlands
13. Physikalisches Institut and Würzburg-Dresden Cluster of Excellence ct.qmat, Universität Würzburg, 97074 Würzburg, Germany
14. Max Planck Institute for Chemical Physics of Solids, 01187 Dresden, Germany
15. Australian Synchrotron
16. Monash University
17. Institut für Theoretische Physik und Astrophysik, Universität Würzburg, 97074 Würzburg, Germany
18. Department of Theoretical Physics, Wrocław University of Science and Technology, Wybrzeże Wyspiańskiego 27, 50-370 Wrocław, Poland
19. Department of Physics and Astronomy, Purdue University, West Lafayette, Indiana 47907 USA
20. School of Physics, University of New South Wales, Sydney 2052, Australia
21. ARC Centre of Excellence in Future Low-Energy Electronics Technologies (FLEET), University of New South Wales, Sydney 2052, Australia
22. Department of Electrical Engineering, Indian Institute of Technology Bombay, Powai, Mumbai 400076, India
23. Institute for Superconducting and Electronic Materials (ISEM), Australian Institute of Innovative Materials (AIIM), University of Wollongong, Wollongong, New South Wales 2525, Australia
24. ARC Centre of Excellence in Future Low-Energy Electronics Technologies (FLEET), University of Wollongong, Wollongong, New South Wales 2525, Australia
25. Würzburg-Dresden Cluster of Excellence ct.qmat, Universität Würzburg, 97074 Würzburg, Germany

*Editors of the roadmap to whom all correspondence should be addressed: b.weber@ntu.edu.sg; michael.fuhrer@monash.edu

**Abstract**

2D topological insulators promise novel approaches towards electronic, spintronic, and quantum device applications. This is owing to unique features of their electronic band structure, in which bulk-boundary correspondences enforces the existence of 1D spin-momentum locked metallic edge states – both helical and chiral – surrounding an electrically insulating bulk. Forty years since the first discoveries of topological phases in condensed matter, the abstract concept of band topology has sprung into realization with several materials now available in which sizable bulk energy gaps – up to a few hundred meV – promise to enable topology for applications even at room-temperature. Further, the possibility of combing 2D TIs in heterostructures with functional materials such as multiferroics, ferromagnets, and superconductors, vastly extends the range of applicability beyond their intrinsic properties. While 2D TIs remain a unique testbed for questions of fundamental condensed matter physics, proposals seek to control the topologically protected bulk or boundary states electrically, or even induce topological phase transitions to engender switching functionality. Induction of superconducting pairing in 2D TIs strives to realize non-Abelian quasiparticles, promising avenues towards fault-tolerant topological quantum computing. This roadmap aims to present a status update of the field, reviewing recent advances and remaining challenges in theoretical understanding, materials synthesis, physical characterization and, ultimately, device perspectives.

**Contents**

# 1. Introduction

Bent Weber[1], Michael S. Fuhrer[2]

[1] Nanyang Technological University, Singapore

[2] School of Physics and Astronomy, and ARC Centre of Excellence in Future Low-Energy Electronics Technologies, Monash University, Monash 3800 Victoria, Australia

**Status**

The year 2022 marks the 40th anniversary of the discovery of the quantum Hall effect (QHE) [1] (Fig. 1) – later understood as the first manifestation of a topologically non-trivial state of matter in 2D [2]. Forty years on, the investigation of topological matter, both theoretically and experimentally, has become one of the most active fields in condensed matter physics. Not only has the field reached a maturity where topologically insulating states are being actively considered for electronic device applications, but 2D topological phases remain a testbed for fundamental questions of condensed matter physics, including the interplay of band topology with electronic correlations and superconductivity towards error resilient quantum information processing.

For nearly one hundred years, the band theory of solids has allowed us to distinguish between conductors and non-conductors – metals and insulators – based on the presence or absence of electronic band gaps. Yet, in the early 1980s, von Klitzing *et al*. [3] showed that a 2D electron gas (2DEG), when subjected to a strong perpendicular magnetic field, would display a quantized Hall resistance $\rho_{x,y} = \frac{h}{\nu e^2}$, arising from metallic states at its boundaries, despite the interior being an insulator. Impossible to explain based on conventional band theory, Thouless [2] put these findings into the context of band topology. Topology in condensed matter describes materials in which electronic band structure is robust to adiabatic (smooth) deformations and that can hence be classified by integer numbers – topological invariants. One of these topological invariants – the Chern number $n \in \mathbb{Z}$ – quantifies the total Berry flux in the Brillouin zone and determines the number of boundary states that emerge toward the trivial vacuum surrounding it. As these boundary modes are a consequence of the quantization of the bulk ("*bulk-boundary correspondence* [4]"), they could be shown to – at least theoretically – pass electrical currents in a dissipationless (lossless) manner.

In 1988, Haldane proposed a toy model for a material in which a non-zero Chern number for a filled band would arise without the requirement of an external magnetic [5]: A hexagonal 2D lattice with antiferromagnetic order would force electrons to move through the lattice at curved trajectories, mimicking the effect of the magnetic field in the quantum Hall state. Such *quantum anomalous Hall* (QAH) insulator would exhibit chiral edge states, similar to case of the QHE. Later, Kane and Mele [6] showed that atomic spin orbit coupling (SOC) can play a similar role in hexagonal lattices such as that of graphene if SOC of the lattice is strong enough to induce a spin-dependent second nearest neighbour hopping which serves to separate spin polarities transverse to a current applied. The net effect can be understood as two identical copies of the Haldane model, one each for each spin polarity, giving rise to a quantized spin Hall conductance at a net zero charge Hall conductance. The *quantum spin Hall* (QSH) state [6] is now the paradigmatic example of the time-reversal invariant 2D topological insulator.

Simultaneously, Bernevig Hughes and Zhang [7] independently proposed a model for the QSH effect based on strain-engineered spin-orbit coupling in semiconductor heterostructures. The QSH state was experimentally detected shortly after [8] in HgTe/CdTe heterostructures (see section 3.1), followed by demonstrations in the related InAs/GaSb system [9]. In both heterostructures, the QSH state arises from exquisite control of quantum well thickness, strain and doping via molecular beam epitaxy (MBE), allowing for a topological bulk gap of 10-20 meV that can easily be resolved at cryogenic temperatures.

The *extrinsic* QSH state in semiconductor heterostructures has the advantage that the inverted quantum well is well-protected as it is embedded within a single-crystalline host matrix. However, its bulk state is also more challenging to probe as it is buried beneath a crystalline epilayer and usually remains inaccessible to surface sensitive probes such as scanning tunnelling microscopy and APRES which have become standard diagnostic tools for the investigation of topological matter (see sections 4.1 and 4.2).

To allow for a wider range of operating parameters in prospective QSH based electronic devices – including high- or even room temperature – a search is ongoing for alternative material system in which the QSH arises as in *intrinsic* property of the atomic lattice, allowing for larger topological band gaps. Given the prohibitively weak spin-orbit coupling in graphene [10], several materials with hexagonal lattices but heavier constituent atoms and strong spin-orbit coupling have been considered such as the 2D Xenes [11] (see section 3.3). By far the largest topological gaps have so far been achieved so far in the bismuthene [12] (see section 3.4) and $Na_3Bi$ [13], ranging into the hundreds of meV. High operating temperatures of up to 100 K have been reported in $WTe_2$ devices [14] (see section 3.2). Beyond QSH materials, 2D topological insulators whose edge states are not susceptible to time-reversal symmetry breaking, are governed by the quantum anomalous Hall (QAH) effect (see section 3.5).

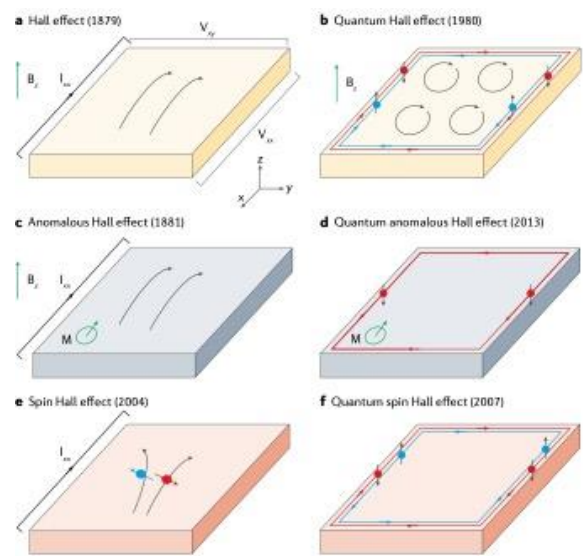

**Figure 1.** Different classes of 2D topological insulators, and their classical counter parts. Reproduced from Ref. [1].

**Current and Future Challenges**

While several promising 2D TI hosts have been identified [8, 11-13, 15], the search for alternative materials platforms continues. Requirements include to achieve large enough topological band gaps to allow 2D TI physics to be demonstrated at room temperature (see sections 5.1 and 5.2), but also versatility to combine 2D TI materials in heterostructures with other functional materials towards spintronics or quantum-electronic applications (e.g. section 2.3). A persistent challenge – across all material systems – remains the optimization of material synthesis protocols to control disorder, crystal size / uniformity, and doping. While the mature molecular beam epitaxy (MBE) techniques applied to semiconductor heterostructure based 2D TIs easily achieve macroscopic uniformity on the wafer scale [8], The introduction of edge roughness disorder especially during device nanofabrication, remains a challenge (see section 3.1). Atomically-thin 2D TIs [12, 13, 15] with their larger topological gaps are intrinsically less prone to disorder. Yet synthesis techniques are not as optimized by far, often giving rise to much higher disorder levels with small grain / domain or island sizes and disordered boundaries of high defect / doping levels (see e.g. section 2.3). While demonstrations of 2D TI physics are manifold, the persistent materials challenges will need to be overcome towards meaningful electronic device applications that go beyond the demonstration of individual prototypes.

Topological insulators present unique opportunities for device applications, but realizing useful topological devices remains challenging. Efforts are ongoing to engineer the 2D TI edge channels (see section 5.1), and seek novel avenues to apply topological materials to realise topological transistors (see section 5.2) and topological electronics (see section 5.3). Approaches towards this goal seek to control the topologically protected bulk or boundary states electrically, or even induce a topological phase transition as a mode to engender switching functionality in prospective devices (see section 5.3). A recent innovative approach is to utilize 2D topological insulators to enhance the functionality of classical field-effect transistors. The proposal aims to overcome "Boltzmann's Tyranny" (see section 5.2) – a fundamental limitations in the rate at which classical transistors can switch (the subthreshold swing) – by exploiting the built-in electric-field sensitivity given by strong Rashba spin-orbit coupling native to many 2D TIs. The field-effect can be further enhanced by integrating negative capacitance materials such as ferroelectrics, promising to overcome Boltzmann's Tyranny and paving the way towards low-power classical electronics.

Finally, 2D TIs may be combined in heterostructures with conventional superconductors such that superconducting pairing may be induced by proximity [16] to realize much-coveted Majorana and other non-Abelian parafermion quasiparticles (see sections 4.2 and 5.1). The versatility of 2D TI materials catalogue for heterostructure engineering, combined with avenues for the tunability of their topological states may ultimately find applications towards error resilient topological quantum information processing [17-19].

**Concluding Remarks**

We note that this roadmap is not meant as a comprehensive review article, but rather as a status update, surveying material systems, materials challenges, and possible device realizations. For excellent reviews and perspectives on 2D TI materials and theory, we refer the reader to the following (non-exhaustive) list of reviews [1, 20-25]. We are aware that several promising materials candidates, characterization techniques, or application aspects may have not been covered, or not been covered to sufficient detail, yet we hope that this roadmap gives an illustrative overview of current progress and challenges in the field.

**Acknowledgements**

The planning of this roadmap project was supported by the National Research Foundation (NRF) Singapore, under the Competitive Research Programme "Towards On-Chip Topological Quantum Devices" (NRF-CRP21-2018-0001) with further support from the Singapore Ministry of Education (MOE) Academic Research Fund Tier 3 grant (MOE2018-T3-1-002) "Geometrical Quantum Materials", and a Singapore MOE AcRF Tier 2 (MOE-T2EP50220-0011). MSF acknowledges support of the ARC Centre of Excellence in Future Low-Energy Electronics Technologies (CE170100039). BW acknowledges a Singapore National Research Foundation (NRF) Fellowship (NRF-NRFF2017-11).

**References**

[1] von Klitzing, K., et al., *40 years of the quantum Hall effect.* Nature Reviews Physics, 2020. **2**(8): p. 397-401.

[2] Thouless, D.J., et al., *Quantized Hall Conductance in a Two-Dimensional Periodic Potential.* Physical Review Letters, 1982. **49**(6): p. 405-408.

[3] Klitzing, K.v., G. Dorda, and M. Pepper, *New Method for High-Accuracy Determination of the Fine-Structure Constant Based on Quantized Hall Resistance.* Physical Review Letters, 1980. **45**(6): p. 494-497.

[4] Hatsugai, Y., *Chern number and edge states in the integer quantum Hall effect.* Physical Review Letters, 1993. **71**(22): p. 3697-3700.

[5] Haldane, F.D.M., *Model for a Quantum Hall Effect without Landau Levels: Condensed-Matter Realization of the "Parity Anomaly".* Physical Review Letters, 1988. **61**(18): p. 2015-2018.

[6] Kane, C.L. and E.J. Mele, *Quantum Spin Hall Effect in Graphene.* Physical Review Letters, 2005. **95**(22): p. 226801.

[7] Bernevig, B.A., T.L. Hughes, and S.-C. Zhang, *Quantum Spin Hall Effect and Topological Phase Transition in HgTe Quantum Wells.* Science, 2006. **314**(5806): p. 1757-1761.

[8] König, M., et al., *Quantum Spin Hall Insulator State in HgTe Quantum Wells.* Science, 2007. **318**(5851): p. 766-770.

[9] Knez, I., R.-R. Du, and G. Sullivan, *Evidence for Helical Edge Modes in Inverted $\mathrm{InAs}/\mathrm{GaSb}$ Quantum Wells.* Physical Review Letters, 2011. **107**(13): p. 136603.

[10] Gmitra, M., et al., *Band-structure topologies of graphene: Spin-orbit coupling effects from first principles.* Physical Review B, 2009. **80**(23): p. 235431.

[11] Bampoulis, P., et al., *Quantum Spin Hall States and Topological Phase Transition in Germanene.* Physical Review Letters, 2023. **130**(19): p. 196401.

[12] Reis, F., et al., *Bismuthene on a SiC substrate: A candidate for a high-temperature quantum spin Hall material.* Science, 2017. **357**(6348): p. 287-290.

[13] Collins, J.L., et al., *Electric-field-tuned topological phase transition in ultrathin Na3Bi.* Nature, 2018. **564**(7736): p. 390-394.

[14] Wu, S., et al., *Observation of the quantum spin Hall effect up to 100 kelvin in a monolayer crystal.* Science, 2018. **359**(6371): p. 76-79.

[15] Fei, Z., et al., *Edge conduction in monolayer WTe2.* Nature Physics, 2017. **13**(7): p. 677-682.

[16] Fu, L. and C.L. Kane, *Superconducting Proximity Effect and Majorana Fermions at the Surface of a Topological Insulator.* Physical Review Letters, 2008. **100**(9): p. 096407.

[17] Mi, S., et al., *Proposal for the detection and braiding of Majorana fermions in a quantum spin Hall insulator.* Physical Review B, 2013. **87**(24): p. 241405.

[18] Alicea, J., *New directions in the pursuit of Majorana fermions in solid state systems.* Reports on Progress in Physics, 2012. **75**(7): p. 076501.

[19] Beenakker, C.W.J., *Search for Majorana Fermions in Superconductors.* Annual Review of Condensed Matter Physics, 2013. **4**(1): p. 113-136.

[20] Hasan, M.Z. and C.L. Kane, *Colloquium: Topological insulators.* Reviews of Modern Physics, 2010. **82**(4): p. 3045-3067.

[21] Qi, X.-L. and S.-C. Zhang, *The quantum spin Hall effect and topological insulators.* Physics Today, 2010. **63**(1): p. 33-38.

[22] Molle, A., et al., *Buckled two-dimensional Xene sheets.* Nature Materials, 2017. **16**(2): p. 163-169.

[23] Cao, C. and J.-H. Chen, *Quantum Spin Hall Materials.* Advanced Quantum Technologies, 2019. **2**(10): p. 1900026.

[24] Li, Z., Y. Song, and S. Tang, *Quantum spin Hall state in monolayer 1T'-TMDCs.* Journal of Physics: Condensed Matter, 2020. **32**(33): p. 333001.

[25] Lodge, M.S., et al., *Atomically Thin Quantum Spin Hall Insulators.* Advanced Materials, 2021. **33**(22): p. 2008029.

## 2.1 Numerical Methods

Xian-Lei Sheng

School of Physics, Beihang University, China

Shengyuan A. Yang

Research Laboratory for Quantum Materials, Singapore University of Technology and Design, Singapore

**Status**

Numerical methods for studying 2D TIs have been well developed. First-principles calculations based on density functional theory (DFT) were widely applied to search for and to expose properties of real TI materials. Via such an approach, starting from a material's crystal structure, one can obtain its band structure and band wave functions for a chosen set of valence electrons. The topological character can be determined by a postprocessing using the DFT band structure information. This can be done via several approaches, e.g., by band symmetry analysis [1][2], by numerical integral of Berry's connection and curvature, by adiabatic deformation, or by Wilson loop method [3][4]. The result is usually reliable, since as a topological invariant, it is intrinsically robust against numerical errors. The topological edge modes for 2D TIs can be studied by calculation on a ribbon-like sample geometry. For small sample size, this is feasible by direct DFT calculation. A widely used approach to speed up this task is to first construct an *ab initio* tight-binding model based on localized Wannier functions [5]. Then, using this model, one can either simulate a finite sample or calculate the edge projected spectrum by an iterative Green's function method. Such *ab initio* tight-binding models can also be used for simulating quantum transport in some device setup. In studying generic physical properties of 2D TIs, one often makes use of low-energy effective models. Such models can be constructed from symmetry constraints with the band symmetry information extracted from DFT calculations. If the low-energy states (e.g., bulk band edges) are confined in a small region of Brillouin zone, one can construct a k·p type model for the description, such as the case for the HgTe quantum wells [6]. If the low-energy states are extended or there are multiple low-energy regions (valleys) that need to be treated simultaneously, then a tight-binding model could be a better choice, such as Kane-Mele model [7]. The above-mentioned numerical methods have been very successful in predicting 2D TI materials and in explaining experimental results. They are now the standard tools for research in this field.

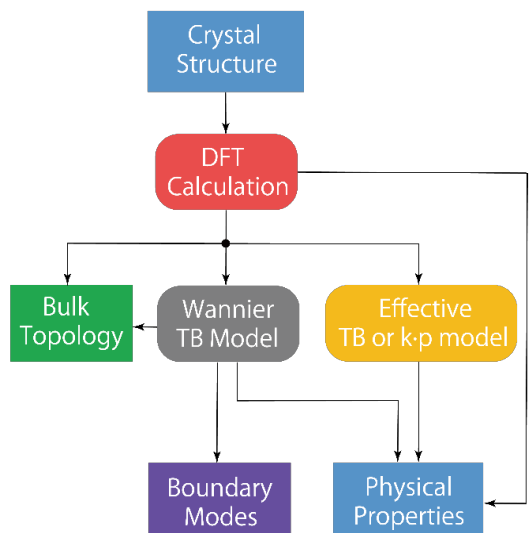

**Figure 1**. Typical workflow of numerical study of topological materials.

**Current and Future Challenges**

Regarding numerical methods for 2D TIs, the challenges come from the following aspects. The first is on accurately determining the bulk band gap values of TIs. Certainly, this band gap problem is a long standing issue intrinsic to density functional theory, not just for TIs. However, the accuracy of gap value is quite important in the current context. For example, the gap of 2D TI material 1T'-WTe$_2$ would be found vanishing if using the generalized gradient approximation (GGA) for exchange-correlation functional (or plus $G_0W_0$ approximation) [8], where the inaccuracy impacts the phase determination. In addition, since most 2D TIs are narrow gap semiconductors, although the accurate gap size does not change the topological character, it does strongly affect the localization of edge modes and the response to external fields. This issue could be especially important for determining the critical point of topological phase transition in 2D TIs where the energy gap closes. Second, since most proposed applications of 2D TIs are based on topological edge modes, we need more accurate approaches to study them in realistic systems. Although the edge modes were expected to be robust transport channels, experiments showed that the robustness can only be maintained at limited channel length or temperature [9]. To fully understand and utilize their physical (especially transport) property, we have to go beyond the current idealized modelling for edge modes and find ways to capture their interaction with defects, impurities, edge re-constructions, inelastic/magnetic scatterings, sample geometry, and etc. Third, devices based on 2D TIs typically require supporting substrate and protecting layers, so we need efficient ways to capture the interaction between 2D TI and these nearby layers. Such interaction is important when the system gap is small, and the conventional computational approach becomes highly expensive when the lattice mismatch results in a large supercell, e.g., in the case of moire layered structures. Finally, we need efficient approaches to handle many-body correlation effects in real materials, since they be could important in the 1D edge channels and affect transport. This is discussed in section 5.1 below.

**Advances in Science and Technology to Meet Challenges**

To meet the above challenges, it requires advances in at least three directions. The first is to deepen our understanding of the fundamental physics of 2D TI systems. For example, to efficiently predict the transport properties of 2D TIs, we need to better understand how the edge channels are affected by various mechanisms and to identify the dominant factors in different regimes. The second is the development of new algorithms that can improve accuracy and meanwhile reduce computational cost. For example, the band gap values can be improved by using meta-GGA or hybrid functional methods, but the computational cost is still quite high, making it difficult to apply such methods to study systems with large cells. Especially, it is challenging when studying large-size heterostructures or moire layered structures. Given that typical device size is on the micron scale, which is $10^4$ above the atomic scale, new multiscale modelling approaches might be promising for investigating certain physical properties, e.g., the edge channel transport. There, the effects of defects, impurities, atomic vibrations may be modelled at a lower (microscopic) level and captured by some parameters/terms in an upper level model for studying the transport property. The third is the advance in computational infrastructure. Of course, regarding numerical studies, powerful computers will enable us to improve calculation accuracy, to adopt more sophisticated approach, and to study larger systems, which are beneficial for addressing all challenges we discussed above.

**Concluding Remarks**

In conclusion, numerical methods for studying 2D TIs have been well developed at current stage. For very small (such as small clusters) or very large (such as ideal bulk periodic systems), current methods can typically result in good accuracy. However, for devices of 2D TIs with sizes (on the order of microns) sitting in between, there are still gaps to be filled in order to efficiently produce reliable results, especially regarding energy gap values, edge transport properties, substrate and correlation effects. In the future, the development of deeper understanding of fundamental physics, new algorithms and approaches, and computation infrastructures will be promising directions to help us address these challenges.

**Acknowledgements**

The authors thank D. L. Deng for valuable discussions. This work was supported by Singapore MOE AcRF Tier 2 (Grant No. MOE-T2EP50220-0011), National Key R&D Program of China (No. 2022YFA1402600) and NSFC (Grants No. 12174018).

**References**

[1] Po, H. C., Vishwanath, A., & Watanabe, H. (2017). Symmetry-based indicators of band topology in the 230 space groups. Nat. Commun., 8, 50. https://doi.org/10.1038/s41467-017-00133-2

[2] Bradlyn, B., Elcoro, L., Cano, J. et al. (2017). Topological quantum chemistry. Nature 547, 298–305. https://doi.org/10.1038/nature23268

[3] Soluyanov, A. A., & Vanderbilt, D. (2011). Wannier representation of $Z_2$ topological insulators. Physical Review B, 83(3), 035108. http://dx.doi.org/10.1103/PhysRevB.83.035108

[4] Yu, R., Qi, X. L., Bernevig, A., Fang, Z., & Dai, X. (2011). Equivalent expression of $Z_2$ topological invariant for band insulators using the non-Abelian Berry connection. Physical Review B, 84(7), 075119. http://dx.doi.org/10.1103/PhysRevB.84.075119

[5] Marzari, N., Mostofi, A. A., Yates, J. R., Souza, I., & Vanderbilt, D. (2012). Maximally localized Wannier functions: Theory and applications. Rev. Mod. Phys. 84, 1419. https://doi.org/10.1103/RevModPhys.84.1419

[6] Bernevig, B. A., Hughes, T. L., & Zhang, S. C. (2006). Quantum spin Hall effect and topological phase transition in HgTe quantum wells. science, 314(5806), 1757-1761. https://doi.org/10.1126/science.1133734

[7] Kane, C. L., & Mele, E. J. (2005). Quantum spin Hall effect in graphene. Physical review letters, 95(22), 226801. https://doi.org/10.1103/PhysRevLett.95.226801

[8] Zheng, F., Cai, C., Ge, S., Zhang, X., Liu, X., Lu, H., Zhang, Y., Qiu, J., Taniguchi, T., Watanabe, K., Jia, S., Qi, J., Chen, J.-H., Sun, D. and Feng, J. (2016), On the Quantum Spin Hall Gap of Monolayer 1T'-WTe2. Adv. Mater., 28: 4845-4851. https://doi.org/10.1002/adma.201600100

[9] Wu, S., Fatemi, V., Gibson, Q. D., Watanabe, K., Taniguchi, T., Cava, R. J., & Jarillo-Herrero, P. (2018). Observation of the quantum spin Hall effect up to 100 kelvin in a monolayer crystal. Science, 359(6371), 76-79. https://doi.org/10.1126/science.abg9094

## 2.2 Optimization principles for Quantum Spin Hall Materials

Ronny Thomale, Institut für Theoretische Physik und Astrophysik and Würzburg-Dresden Cluster of Excellence ct.qmat, Julius-Maximilians-Universität Würzburg, Germany

**Status**

From its theoretical conception and first experimental discovery, the quantum spin Hall effect (QSHE) took the condensed matter community by storm. After predecessors such as superfulid Helium-3 and the quantum Hall effect, the QSHE truly put the field of topological quantum phases on the map, and has ever since been the center of innovation and progression within topological matter.
As much as the principal phenomenology of QSHE by now is well understood, the microscopic mechanisms leading up to it continue to diversify [Fig. 1]. The original Kane-Mele mechanism [1] proposes a Dirac cone gap opening due to atomic spin-orbit coupling while the Bernevig-Hughes-Zhang mechanism [2] starts from a point group symmetry reduction imposed by a semiconductor quantum well combined with band inversion. While the latter was immediately realized in HgTe/CdTe quantum wells [3], it took a systematic material analysis to eventually identify the candidate material jacutingaite with a large Kane-Mele gap [4]. Spectroscopically, the record high-temperature QSHE to date has been set by bismuthene, a heterostructure of a bismuth monolayer and a SiC substrate [5]. The key insight behind this advancement is an improvement of the original Kane-Mele proposal: Instead of only allowing the spin-orbit-mediated gap opening to enter as a higher order process, the Bi/SiC low energy structure features a px/py two-orbital manifold, where the px and py orbital are coupled to each other via spin-orbit coupling in first order [6]. This leads to a Dirac gap roughly twice of the atomic spin orbit coupling scale of Bi within density functional theory calculations, and is even exceeded by the measured gap to be as high as 800 meV. Shortly thereafter, a record temperature of QSH transport was found for a WTe2 monolayer [7]. Despite a moderate bulk spectral gap, the transport gap appears to be of a size enabling the observation of quantized QSH edge channels up to 100 kelvin. This can be reconciled from a particular feature of electronic bulk Dirac cones in WTe2 named custodial glide symmetry [8], where the gap opening of the Dirac cones located away from high symmetry momenta is moderate yet the direct gap within the momentum regime supported by the QSHE edge modes is large.

**Current and Future Challenges**

From an experimental perspective, one of the biggest challenges lies in the preparation of gated QSH candidate samples for transport measurements and applications.
Transport experiments proved somewhat feasible in HgTe/CdTe and related quantum wells [3,9] due to their outstandingly pure synthesis through advanced growth techniques. In bismuthene [5], however, domain structure, gating, and stability are the central current bottlenecks. It is even still challenging to compose a suitable capping layer, in order to facilitate sample proliferation to ex situ environments. For jacutingaite, its multi-atomic composition poses a challenge for sample preparation. While oligolayers have successfully been synthesized, the monolayer limit is still out of reach. As much as the edge modes in ideal WTe2 samples appear highly suited to high-temperature QSH transport [7], the sample growth is nearly as challenging as in bismuthene. Given its potentially custodial glide symmetry character, this might be related to a high sensitivity towards the specific armchair-zigzag termination of the respective WTe2 sample.
From a theoretical perspective, the band structure search for quantum spin Hall candidate materials could rather directly be executed for Kane-Mele type materials, but not for other QSHE classes. This particularly applies to non-universal impact such as point group symmetry breaking imposed by quantum wells or orbital filtering imposed by heterostructures, which are quintessential to the QSHE formation in HgTe/CdTe and bismuthene. In addition to finding the appropriate kinematic modelling,

QSHE candidates such as WTe2 pose even more intricate questions about interacting topological insulators. The d orbitals of W are responsible for sizable electronic correlations, giving rise to multifold many-body physics such as excitonic insulator behaviour. To which extent electronic correlations influence the bulk QSH character of WTe2 is subject of ongoing research. In general, interactions could help in reducing the low-energy density of states to an advantageous setting for QSHE behaviour, potentially complementing the orbital filtering mechanism of the substrate component of several QSHE heterostructures. It is likely that interactions will become a tuning knob of the future in order to single out the ideal QSHE candidate material.

**Advances in Science and Technology to Meet Challenges**

QSHE samples could serve as an efficient spin splitter for spintronics applications. In particular, QSHE edge modes could yield nearly dissipationless quantum channels for low-power electronics without an external magnetic field. Even though quantum Hall effect had long been realized at high temperature in technologically convenient compounds such as graphene, the need for an external magnetic field is the key property which has prevented the use of quantum Hall edge modes for next-generation low power electronics. Ideally, the QSHE could combine the advantages of quantum Hall edge modes with the absence of an external field. QSHE edge modes can in principle exhibit similar, but not identical dissipationless conduction properties as quantum Hall edge modes. This is because even though elastic backscattering is forbidden by time reversal symmetry, inelastic backscattering of counterpropagating QSHE edge modes is unavoidable. Furthermore, given the plethora of potential QSHE candidates at elevated temperatures, phonons might be sufficiently activated to generate additional dissipation channels. Improved crystal growth of existing QSHE candidates along with the theoretical and experimental search for better candidate materials are expected to improve the status of QSHE applicability to technological design. In case better sample growth can be accomplished, bismuthene promises to be most suited for applications not only due to its record high operational temperature, but also due to the mass availability of its constituents Bi and SiC. Further advances to improve the status quo of QSHE are likely to derive from identifying additional QSHE formation mechanisms. For instance, a triangular lattice of indium, labelled indenene, has recently been suggested as a versatile mono-atomic compound realization of QSHE [10]. There, the Wannier centers of the effective hybrid orbits are not located on the atomic sites, but at the midpoint between them. Departing from a triangular lattice, this yields an effective honeycomb lattice model located in the topologically non-trivial QSHE regime. The growth properties of indenene appear advantageous as compared to competing material candidates. Even though the compound itself in the end might not turn out to be the ideal QSHE setting [10], it underlines the continued revelation of novel QSHE formation principles, and broadens the scope in which the ideal QSHE material can be explored.

**Concluding remarks**

As much as the field of topological matter, and topological insulators in particular, has triggered excitement and activity in fundamental sciences over the past decades, a pathway towards enabling a unique technical advancement or functionality is still rather unclear. Quantum spin Hall effect (QSHE) appears as the most concise motif of a topological quantum phase which may unfold in a technological context. The variety of QSHE materials allow for a large landscape to be explored and optimized in the future, and promise a gold standard for topological functionality in correlated electron systems.

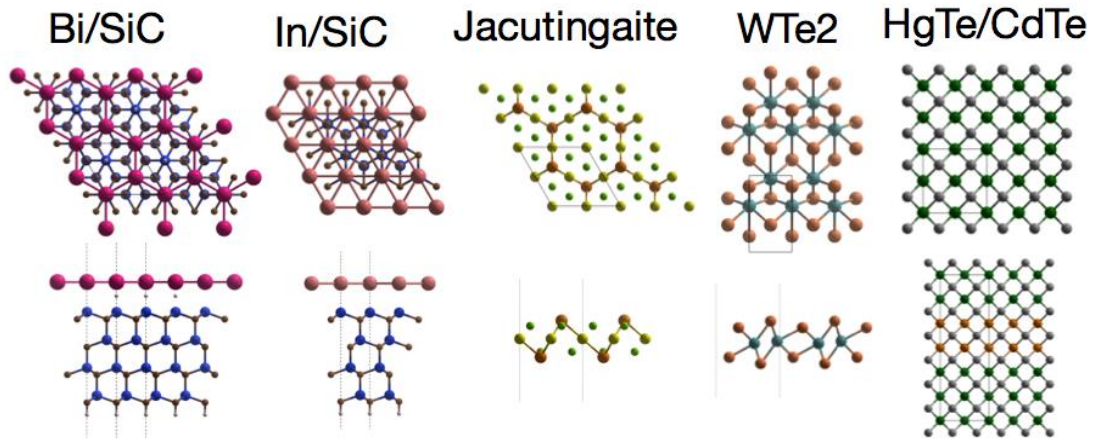

Fig. 1. Synopsis of pradigmatic quantum spin Hall (QSH) material candidates: Bismuthene Bi/SiC, a heterostructure of SiC and a Bi monolayer so far sets the record qsh spectral gap [5,6]. Indenene In/SiC features a triangular monolayer on top of SiC and still forms an effective honeycomb lattice for its Wannier centers [10]. Jacutingaite is the most promising candidate for the QSH formation mechanism traced back to the original proposal for graphene [1,4]. Tungsten ditelluride WTe2 exhibits additional qsh stability features due to a custodial glide symmetry [7,8]. For the first realization of qshe in HgTe/CdTe [2,3], the quantum well geometry enforces a point group symmetry reduction to transform a semiconductor into a topological insulator [3,9].

**Acknowledgments**

RT is funded by the Deutsche Forschungsgemeinschaft (DFG, German Research Foundation) through Project-ID 258499086 - SFB 1170,the W\"{u}rzburg-Dresden Cluster of Excellence on Complexity and Topology in Quantum Matter-ct.qmat Project-ID 390858490 - EXC 2147, and FOR QUAST 5249-449872909 (Project P3).

**References**

[1] C. L. Kane and E. J. Mele, "Quantum Spin Hall Effect in Graphene", Phys. Rev. Lett. 95, 226801 (2005).

[2] B. A. Bernevig, T. L. Hughes, and S.-C. Zhang, "Quantum Spin Hall Effect and Topological Phase Transition in HgTe Quantum Wells", Science 314, 1757 (2006).

[3] M. König, S. Wiedmann, C. Brüne, A. Roth, H. Buhmann, L. W. Molenkamp, X.-L. Qi, and S.-C. Zhang, "Quantum Spin Hall Insulator State in HgTe Quantum Wells", Science 318, 766 (2007).

[4] A. Marrazzo, M. Gibertini, D. Campi, N. Mounet, and N. Marzari, "Prediction of a Large-Gap and Switchable Kane-Mele Quantum Spin Hall Insualtor", Phys. Rev. Lett. 120, 117701 (2018).

[5] F. Reis, G. Li, L. Dudy, M. Bauernfeind, S. Glass, W. Hanke, R. Thomale, J. Schäfer, and R. Claessen, "Bismuthene on a SiC substrate: A candidate for a high-temperature quantum spin Hall material", Science 357, 287 (2017).

[6] G. Li, W. Hanke, E. M. Hankiewicz, F. Reis, J. Schäfer, R. Claessen, C. Wu, and R. Thomale, "Theoretical paradigm for the quantum spin Hall effect at high temperatures", Phys. Rev. B 98, 165146 (2018).

[7] S. Wu, V. Fatemi, Q. D. Gibson, K. Watanabe, T. Taniguchi, R. J. Cava, and P. Jarillo-Herrero, "Observation of the quantum spin Hall effect up to 100 kelvin in a monolayer crystal", Science 359, 76 (2018).

[8] S. Ok, L. Muechler, D. Di Sante, G. Sangiovanni, R. Thomale, and T. Neupert, "Custodial glide symmetry of quantum spin Hall edge modes in monolayer WTe2", Phys. Rev. B 99, 121105(R), 2019.

[9] I. Knez, R.-R. Du, and G. Sullivan, "Evidence for Helical Edge Modes in Inverted InAs / GaSb Quantum Wells", Phys. Rev. Lett. 107, 136603 (2011).

[10] M. Bauernfeind, J. Erhardt, P. Eck, P. Thakur, J. Gabel, T.-L. Lee, J. Schäfer, S. Moser, D. Di Sante, R. Claessen, and G. Sangiovanni, "Design and realization of topological Dirac fermions on a triangular lattice", Nature Communications 12, 5396 (2021).

### 3.1 Two-dimensional topological insulators in semiconductor heterostructures

*Saquib Shamim[1,2,3] and Laurens W. Molenkamp[1,2]*
[1]Experimentelle Physik III, Physikalisches Institut, Universität Würzburg, Am Hubland, 97074, Würzburg, Germany
[2]Institute for Topological Insulators, Universität Würzburg, Am Hubland, 97074, Würzburg, Germany
[3]Department of Condensed Matter and Material Physics, S. N. Bose National Centre for Basic Sciences, Kolkata 700106, India

**Status**

Semiconductor heterostructures have proven to be an invaluable platform to probe many fundamental concepts in physics, such as metal-insulator transition, spin-transport phenomena, and the quantum Hall effect - the first topological state of matter discovered. Two-dimensional topological insulator though first predicted in graphene but was experimentally realized in thin HgTe quantum wells, a narrow gap semiconductor.

Two-dimensional topological insulators show the quantum spin Hall effect, which is characterized by an insulating bulk but with helical edge channels protected under time-reversal symmetry. Motivated by extensive research on the intrinsic spin Hall effect in semiconductors and insulators, two papers predicted the quantum spin Hall effect: for graphene [1] and HgTe/CdTe heterostructures [2], with mutually inverted band structures. The experimental observation in graphene proved difficult due to a small spin-orbit gap [1]. It was more feasible to realize the effect in HgTe quantum wells sandwiched between (Hg,Cd)Te barriers [2,3]. Konig et al., used molecular beam epitaxy to grow the HgTe/(Hg,Cd)Te layers where the thickness of the HgTe layer was tuned to ensure that the band structure remains inverted but with a bulk band gap of around 10-20 meV, which can be easily resolved in transport experiments. The conductance of devices fabricated from such material was quantized to $2e^2/h$ in the absence of any external magnetic field, a hallmark of the quantum spin Hall effect [3]. Further experimental evidence, such as nonlocal transport, scanning squid spectroscopy, and induced superconductivity, have unambiguously demonstrated the existence of the quantum spin Hall effect.

The main advantage of using semiconductor heterostructures, particularly HgTe/(Hg,Cd)Te for quantum spin Hall studies lies in the high crystal quality of the material grown by molecular beam epitaxy, which results in mobilities of more than a million (for HgTe quantum wells). The excellent control of thickness allows for studying the phase transition from topological to a trivial insulator. The band structure can be engineered by varying the strain to realize a large band gap of up to 55 meV [4]. These heterostructures can be alloyed with other magnetic atoms and superconductors to explore various aspects of topological superconductivity. Thus, this research field is significant for aspects of fundamental physics as well as applications. However, certain challenges need to be addressed for timely and efficient progress in this direction, as discussed below.

**Current and Future Challenges**

One of the major issues which plague experimental research on narrow-gap semiconductors, including HgTe-based two-dimensional topological insulators, is the occurrence of charge puddles in these systems. These charge puddles are due to an inhomogeneous potential landscape formed because of growth defects, vacancies, fabrication processes, etc. When the electrochemical potential is tuned into the bulk band gap to access the quantum spin Hall states, small charged regions are formed (shown by circles in Fig. 1a), that contain bulk carriers. These charged regions have a profound effect on the measured properties of the quantum spin Hall states even in the bulk band gap as elaborated below.

Though the quantum spin Hall effect has been successfully and unambiguously demonstrated in various material systems, the conductance quantization observed experimentally in the quantum spin Hall effect is nowhere close to that observed in the quantum Hall effect, where the conductance is perfectly quantized to multiples of $e^2/h$. In contrast, all demonstrations of the quantum spin Hall effect such as in HgTe and $WTe_2$ show reproducible fluctuations in conductance [3,5]. The magnitude of these fluctuations varies for different systems and even among various devices fabricated from the same material system and can be as large as a few kΩ. Theoretically, these fluctuations have been ascribed to inelastic scattering due to charge puddles [6,7]. It is an experimental challenge to realize devices where the conductance is quantized to $2e^2/h$ without any fluctuations.

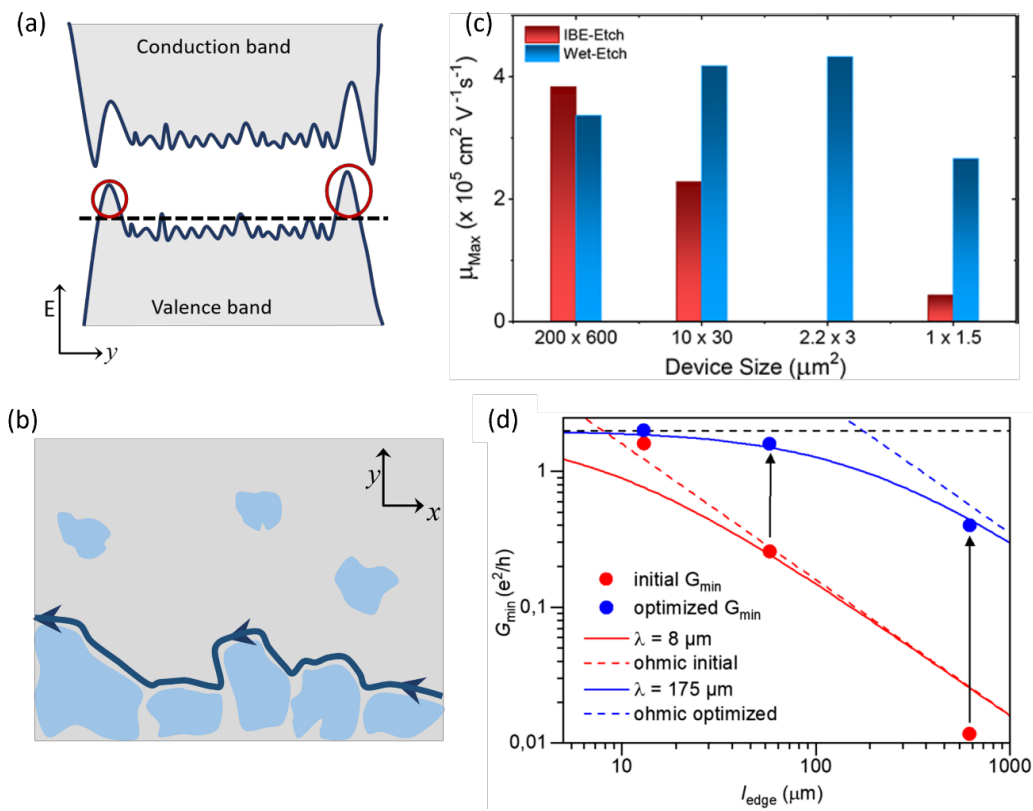

**Figure 1.** (a) A schematic of the potential landscape showing the formation of charge puddles at the edge of the device. E is the energy and y is along the width of the device. (b) Schematic showing a quantum Hall edge channel (blue line) along a network of charge puddles created by the potential landscape. x is along the length of the device. (a) and (b) are adapted from Ref. [11]. Copyright 2022, The Authors. (c) A bar graph showing a comparison of the mobility of devices of various dimensions fabricated using the dry-etch process (red bars) and the wet-etch process (blue bars). Reprinted with permission from Ref. [12]. Copyright (2018), American Chemical Society. (d) The measured conductance within the bulk band gap ($G_{min}$) as a function of channel length for as-fabricated devices (red dots) and after gate-training (blue dots). The black dashed line shows the expected conductance of $2e^2/h$. The solid lines indicated fits from which average scattering length $\lambda$ is extracted. The dashed lines correspond to ohmic behaviour where $G_{min} \propto 1/l_{edge}$. Reprinted with permission from Ref. [8]. Copyright (2019) by the American Physical Society.

The conductance quantization due to the quantum spin Hall effect is restricted to microscopic devices of dimensions ranging from 100s of nm to a few microns depending on the material system [5,8]. It has been repeatedly shown that for devices of larger dimensions, the measured conductance in the quantum spin Hall regime is always less than $2e^2/h$ [3,8]. The larger the devices, the more the suppression of the conductance from the expected value [8]. Apart from being a fundamental problem in the investigation of various aspects of the quantum spin Hall effect, this also poses limitations for applications of quantum spin Hall devices. Such suppression of conductance from the expected quantized value has been attributed to scattering from charge puddles [8]. Microwave spectroscopy measurements in the bulk band gap show that the measured density of states is more than 20 times larger than that expected from a helical edge state [9]. The additional contribution is due to charge puddles which couple with the edge channels leading to scattering.

Another serious consequence of charge puddles is that they lead to the observation of conductance in the regime where one expects a trivial insulating state [10,11]. When a 2D TI is subjected to an external magnetic field, the inversion of Landau levels is lifted and the system transitions to a trivial insulating state where the expected conductance is zero [11]. However, electrical conductance and microwave impedance spectroscopy measurements revealed an expected edge conductance in the insulating regime [10]. This has now been identified as due to fabrication-induced damage to the material, due to which a quantum Hall edge channel propagates along a network of charge puddles as demonstrated in Ref. [11] (Fig. 1b). Finally, one of the challenges for the research on the quantum spin Hall effect is the scarcity of the number of model systems where the conductance quantization has been experimentally demonstrated.

**Advances in Science and Technology to Meet Challenges**

This section discusses recent advances in science and technology that have partially addressed some of the challenges mentioned above. One of the prerequisites to experimentally investigate electron transport in quantum spin Hall effect is to realize pristine edges with fully transmitting edge channels. Conventional dry-etching, particularly for meso- and nanoscale devices, leads to damaged edges. To realize clean edges, the first advancement is in the fabrication process of meso-and nanoscale devices using chemical wet-etching as opposed to conventional dry-etching [12]. The wet-etching process results in microstructures with mobility comparable to macroscopic devices (Fig. 1c) and quantized spin Hall conductance in devices as long as 10 $\mu$m [12]. Using wet-etching, clean one-dimensional quantum point contacts were fabricated in HgTe based 2D topological insulator and the interaction between the quantum spin Hall edge channels has been investigated [13]. Further, side-contacted Josephson junctions in 2D topological insulators have been realized with this technique to explore concepts of topological superconductivity [12]. Further, wet-etched devices show the theoretically expected trivial insulating state when the inversion of Landau levels is lifted under an external magnetic field, clearly establishing that previous observation of edge conduction in this regime is due to disorder [11].

The next advancement has been to electrostatically modify the potential landscape of the charge puddles to mitigate its effects on the transport properties of the quantum spin Hall edge channels [8]. This is achieved via continuous charging and discharging of puddles by using an electric field, which smoothens the potential landscape and results in an increase in the conductance of the macroscopic devices to approach closer to the expected value of $2e^2/h$ (Fig. 1d).

In spite of the above advancements, further refinement in both growth and device fabrication is essential such that the quantum spin Hall devices with clean and fully transmitting edge channels can be realized. In terms of growth, for existing materials, it is essential to have excellent crystal quality with as few defects as possible. Additionally, the growth of different model systems, where the quantum spin Hall effect has been predicted, will further boost experimental research in this field. An exciting way to probe the pristine quantum spin Hall edge channels is to grow the materials and in-situ perform the cross-sectional scanning tunneling microscopy on the as-grown materials, as has already been done for (Pb,Sn)Se [14] and Bi [15]. However, for device applications, various fabrication processes are unavoidable. Novel lithographic and etching techniques are required to eliminate the effects of charge puddles, which would be quite challenging.

**Concluding Remarks**

To conclude, we wish to emphasize that quantum wells in HgTe/(Cd,Hg)Te have provided an excellent platform to investigate the quantum spin Hall effect. Novel etching techniques and lithographic control have reduced the influence of charge puddles on transport in the quantum spin Hall regime. Further lithographic techniques, like STM lithography, may be pursued to explore if charge puddle-free quantum spin Hall structures can be realized. For electrical transport experiments, samples without bulk doping are absolutely necessary, whilst for STM and ARPES measurements, doping can be beneficial. Finally, the limited number of model systems where the quantum spin Hall effect has been experimentally demonstrated limits the experimental research on this topic. In the last few years, more systems have been predicted to host quantum spin Hall states and it would be remarkable if experimental demonstrations for the same are reported such that there is a worldwide concentrated effort to address the challenges in the field.

**Acknowledgements**

We acknowledge financial support from the Deutsche Forschungsgemeinschaft (DFG, German Research Foundation) in the Leibniz Program; in the projects SFB 1170 (Project ID 258499086) and from the Würzburg-Dresden Cluster of Excellence on Complexity and Topology in Quantum Matter (EXC 2147, Project ID 39085490), and the Institute for Topological Insulators. S.S. thanks S. N. Bose National Centre for Basic Sciences, for financial support.

**References**

1) C. L. Kane and E. J. Mele, Quantum spin Hall effect in graphene, *Phys. Rev. Lett.* **95**, 226801 (2005)
2) B. A. Bernevig, T. L. Hughes, and S.-C. Zhang, Quantum spin Hall effect and topological phase transition in HgTe quantum wells, *Science* **314**, 1757 (2006)
3) M. König, S. Wiedmann, C. Brüne, A. Roth, H. Buhmann, L. W. Molenkamp, X.-L. Qi and S.-C. Zhang, Quantum spin Hall insulator state in HgTe quantum wells, *Science* **318**, 766 (2007)
4) P. Leubner, L. Lunczer, C. Brüne, H. Buhmann, and L. W. Molenkamp. Strain Engineering of the Band Gap of HgTe Quantum Wells Using Superlattice Virtual Substrates. *Phys. Rev. Lett.* **117**, 086403 (2016)
5) S. Wu, V. Fatemi, Q. D. Gibson, K. Watanabe, T. Taniguchi, R. J. Cava, and P. J.-Herrero. Observation of the quantum spin Hall effect up to 100 kelvin in a monolayer crystal. *Science* **359**, 76 (2018).
6) J. I. Väyrynen, M. Goldstein, and L. I. Glazman. Helical Edge Resistance Introduced by Charge Puddles. *Phys. Rev. Lett*. **110**, 216402 (2013)
7) J. I. Väyrynen, M. Goldstein, Y. Gefen, and L. I. Glazman. Resistance of helical edges formed in a semiconductor heterostructure. *Phys. Rev. B* **90**, 115309 (2014)
8) L. Lunczer, P. Leubner, M. Endres, V. L. Müller, C. Brüne, H. Buhmann, and L. W. Molenkamp, Approaching quantization in macroscopic quantum spin Hall devices through gate training, *Phys. Rev. Lett*. **123**, 047701 (2019)
9) M. C. Dartiailh, S. Hartinger, A. Gourmelon, K. Bendias, H. Bartolomei, H. Kamata, J.-M. Berroir, G. Fève, B. Plaçais, L. Lunczer, R. Schlereth, H. Buhmann, L. W. Molenkamp, and E. Bocquillon. Dynamical Separation of Bulk and Edge Transport in HgTe-Based 2D Topological Insulators. *Phys. Rev. Lett*. **124**, 076802 (2020)
10) E. Y. Ma, M. R. Calvo, J. Wang, B. Lian, M. Mühlbauer, C. Brüne, Y.-T. Cui, K. Lai, W. Kundhikanjana, Y. Yang, M. Baenninger, M. König, C. Ames, H. Buhmann, P. Leubner, L. W. Molenkamp, S.-C. Zhang, D.

Goldhaber-Gordon, M. A. Kelly and Z.-X. Shen. Unexpected edge conduction in mercury telluride quantum wells under broken time-reversal symmetry. *Nat. Commun*. **6**, 7252 (2015).

11) S. Shamim, P. Shekhar, W. Beugeling, J. Böttcher, A. Budewitz, J.-B. Mayer, L. Lunczer, E. M. Hankiewicz, H. Buhmann, and L. W. Molenkamp. Counterpropagating topological and quantum Hall edge channels. *Nat. Commun.* **13**, 2682 (2022)

12) K. Bendias, S. Shamim, O. Herrmann, A. Budewitz, P. Shekhar, P. Leubner, J. Kleinlein, E. Bocquillon, H. Buhmann, and L. W. Molenkamp, High mobility HgTe microstructures for quantum spin Hall studies, Nano Lett. 18(8), 4831 (2018)

13) J. Strunz, J. Wiedenmann, C. Fleckenstein, L. Lunczer, W. Beugeling, V. L. Müller, P. Shekhar, N. T. Ziani, S. Shamim, J. Kleinlein, H. Buhmann, B. Trauzettel, and L. W. Molenkamp. Interacting topological edge channels. *Nat. Phys.* **16**, 83 (2020)

14) P. Sessi, D. D. Sante, A. Szczerbakow, F. Glott, S. Wilfert, H. Schmidt, T. Bathon, P. Dziawa, M. Greiter, T. Neupert, G. Sangiovanni, T. Story, R. Thomale, and M. Bode. Robust spin-polarized midgap states at step edges of topological crystalline insulators. *Science* **354**, 1269 (2016)

15) F. Schindler, Z. Wang, M. G. Vergniory, A. M. Cook, A. Murani, S. Sengupta, A. Y. Kasumov, R. Deblock, S. Jeon, I. Drozdov, H. Bouchiat, S. Guéron, A. Yazdani, B. A. Bernevig & T. Neupert. Higher order topology in bismuth. *Nat. Phys.* **14**, 918 (2018)

## 3.2 Tungsten Ditelluride (WTe$_2$)

David Cobden[1], Dmytro Pesin[2]
[1]University of Washington, Seattle    [2]University of Virginia

**Status**

The layered transition metal dichalcogenides (TMDs) can have topologically interesting bands and can be either metallic, semiconducting, or semimetallic, depending on the choice of metal and chalcogen, on the particular structure of each layer, and on how the layers are stacked. Those where the layers have the low-symmetry 1T' structure (Fig. 1a) exhibit band inversion and overlapping momentum-separated conduction and valence bands Figs. 1b,c). An isolated 1T' monolayer is predicted to become a 2DTI if, for whatever reason, the band overlap is absent[1]. Two TMDs are known to be stable in the 1T' state: MoTe$_2$, which is metallic down to the monolayer (1L) limit; and WTe$_2$, which is normally semimetallic and becomes insulating in the monolayer limit with an effective gap of about 50 meV at low temperatures[2] (Fig. b).

Scanning tunneling spectroscopy (STS) on 1L WTe$_2$ (Fig. 1d) shows signs of suppression of the gap at the edges[2], [3] (Fig. 1e), and microwave impedance microscopy (MIM) reveals robust conductivity localized to the edges at helium temperatures[4] (Fig. 1f). Multiterminal resistance measurements demonstrate the presence of an edge contribution as high as room temperature[5], [6], with the edges dominating conduction below ~100 K (Figs. 2a,b). The two-terminal conductance can approach the single-channel limit of $e^2/h$ for very short edges but it decreases with increasing length, implying that backscattering occurs even at zero magnetic field. The conductance is suppressed by a magnetic field oriented perpendicular to a certain axis in the crystal[7] (Fig. 1g) which can be identified with the momentum-independent spin axis inherited from the bulk bands near $E_F$ in a k.p analysis[8].

Monolayer WTe$_2$ thus has all the expected characteristics of a 2DTI. It can be prepared by exfoliation of crystals that are readily grown having high quality, although it must be protected by encapsulation (typically using hBN) to prevent rapid oxidation. It can serve as a model system for studying helical edge modes in their own right and in concert with other electronic phenomena, both because the material itself combines topology with superconductivity and correlation effects (see below) and because it can be placed in intimate contact with other layered materials having different properties.

**Current and Future Challenges**

The most obvious open question, applicable to all 2DTIs and key to potential applications, is what limits the edge conductance at zero magnetic field where single-particle backscattering is prohibited. Most known scattering mechanisms are suppressed as a power of $T$, leading to higher conduction as $T \to 0$. However, in practice the edge conduction freezes out below a few kelvin. The linear-response conductance shows large mesoscopic fluctuations with gate voltage $V_g$, and once it is frozen out, there is an apparent threshold bias for current flow on the order of a hundred μeV that fluctuates with $V_g$ and that increases monotonically with magnetic field (Fig. 2c). Although understanding of the origin of this freeze-out in zero magnetic field is presently lacking, a simple mechanism of magnetoresistance in which electrons flip their spin via Larmor precession on localized states can account for the observed linear and nonlinear magnetotransport at higher temperatures. This observation hints that the zero-field problem may not be intractable.

There are many indications that 1L WTe$_2$ obtains its insulating behavior from correlation effects[9]. It becomes a superconductor[10], [11] (with $T_c$~800 mK) at a modest electron doping of ~$5 \times 10^{12}$ cm$^2$ (Fig. 2d), representing too small a Fermi surface to support superconductivity in an uncorrelated conduction band. Its gating characteristics are starkly incompatible with those of a small-gap semiconductor (Fig. 2e). The exciton binding energy is calculated to be several hundred meV, raising the possibility that the neutral state is an excitonic topological insulator. The transition from topological insulator to superconductor may therefore involve a competition between electron-hole pairing and electron-electron pairing, but theoretical understanding of bulk correlation physics in 1L WTe$_2$ is limited at present. However, a number of arguments have been made that edge states connected to single-particle band topology are likely quite robust against bulk interactions.

Another open question is whether the spin polarization can be transferred in or out of the edge modes, for spintronics applications. Addressing it will require developing weak tunnel contacts to the edges (the present metal contacts are too intrusive) and interfacing with magnetic materials. A related question is whether the edge modes can be made superconducting by proximitization, thus providing a route to Majorana zero modes. This calls for developing reliable contacts with superconducting materials such as $NbSe_2$ flakes or sputtered NbN.

**Advances in Science and Technology to Meet Challenges**

The biggest obstacles to understanding and exploiting the 2DTI properties of 1L $WTe_2$ lie in device fabrication. For most purposes, a large enough monolayer flake needs to be identified and characterized and then incorporated into a van der Waals stack using polymer stamps, during which process it can easily oxidize and crack. Air sensitivity makes it hard to pattern the flakes into useful shapes such as Hall bars or to control the geometry and termination of edges, and degrades the quality of electrical contacts. These problems can probably be overcome with concerted effort, for instance by establishing etching and patterning techniques compatible with glove boxes, using different substrates, stamps and metals, adjusting crystal growth parameters, doping contacts, and so on. With samples of optimal geometry that are free of damage, and good contacts that can be chosen to be of various kinds, many more experimental techniques can be methodically brought to bear. Consequences of imperfections, edge geometry and termination, etc., can then be established and new directions explored, for example employing both low-ohmic and weak-tunneling contacts with superconducting and ferromagnetic materials. Techniques are being introduced to measure thermal and thermoelectric coefficients of 2D materials which can probe electron-hole correlations in the bulk, and apparatus is being developed to allow applying uniaxial strain which could induce a topological transition. There are many possibilities on the crystal growth side too, such as incorporating magnetic dopants with the possibility of suppressing the edge states or of developing a quantum anomalous Hall state, and exploring other related layered tellurides with similar topological band structure, such as $TaIrTe_4$.

**Concluding Remarks**

Monolayer $WTe_2$ can be prepared by exfoliation and stacking or by molecular beam epitaxy and may be the most convenient 2DTI currently available, despite the facts that it degrades in air and there is as yet no method for obtaining single-crystal sheets on the mm scale. Thanks to a large bulk gap it exhibits helical edge transport even at liquid nitrogen temperature, and, surprisingly, regardless of the low structural symmetry at the edges, the spin polarization is along a unique, well defined axis. These factors are highly advantageous for electronic control and manipulation. On the other hand, the edge channels have substantial resistivity of uncertain origin and the bulk gap appears to be of a many-body nature, presenting open challenges to theory and modelling. In any case, monolayer $WTe_2$ provides a promising system for exploring the physics of 2DTIs, and there are clear paths ahead for improving device properties.

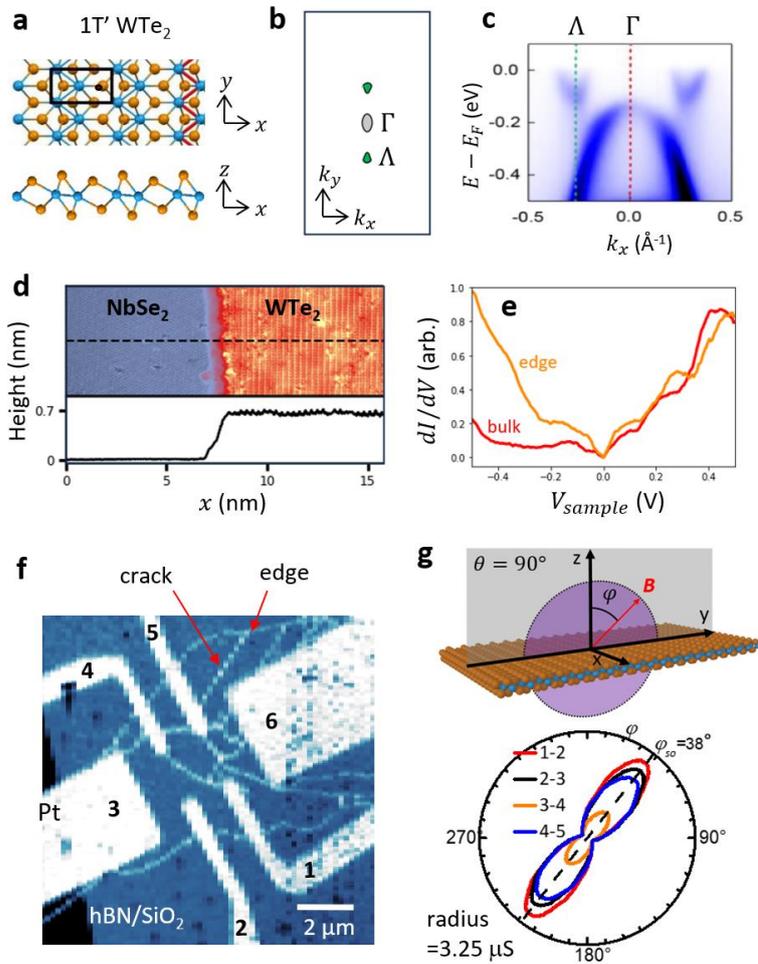

Figure 1. **a**, Structure of monolayer (1L) WTe$_2$. **b**, Brillouin zone indicating electron (green, at $\pm\Gamma$) and hole (gray, at $\Lambda$) pockets. **c**, ARPES spectrum of K-doped 1L WTe$_2$ on graphite [2]. **d**, STM image and line trace of 1L WTe$_2$ on a NbSe$_2$ substrate, and **e**, STS spectrum near the edge and in the bulk [3]. **f**, MIM image of local conductivity in a 1L WTe$_2$ device [4]. **g**, Polar plot of conductance vs angle $\phi$ in the mirror plane of the WTe$_2$ strcture for different pairs of contacts on the device in **f**.

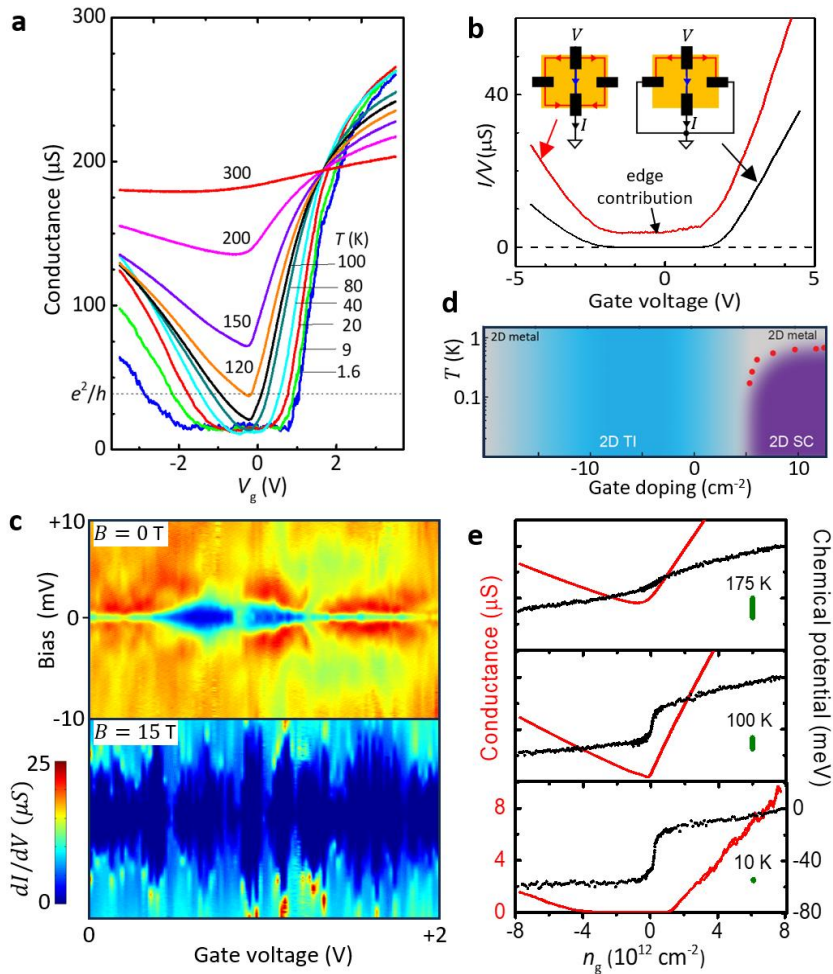

Figure 2. **a,** Gate and temperature dependence of the conductance between adjacent contacts on a 1L WTe$_2$ flake encapsulated in hBN. **b**, Exclusion of edge conduction by partial current measurement, as shown in the insets. **c**, Differential conductance vs bias and gate voltage at zero magnetic field and at 15 T. **d**, Schematic phase diagram of 1L WTe$_2$. **e**, Anomalous behavior of bulk conductance and chemical potential near zero doping suggests the presence of excitons in the equilibrium state and possible excitonic insulator forming below 100 K.

**Acknowledgements**

D.C. was supported by Programmable Quantum Materials, an Energy Frontier Research Center funded by the U.S. Department of Energy (DOE), Office of Science, Basic Energy Sciences (BES), award DE-SC0019443, and by NSF MRSEC Grant No. 1719797.  D.P. was supported by NSF Grant No. DMR-2138008.

**References**

[1]　X. Qian, J. Liu, L. Fu, and J. Li, "Quantum spin Hall effect in two-dimensional transition metal dichalcogenides," *Science*, vol. 346, no. 6215, pp. 1344–1347, Dec. 2014, doi: 10.1126/science.1256815.
[2]　S. Tang *et al.*, "Quantum spin Hall state in monolayer 1T'-WTe$_2$," *Nature Physics*, vol. 13, no. 7, Art. no. 7, Jul. 2017, doi: 10.1038/nphys4174.

[3]  F. Lüpke *et al.*, "Proximity-induced superconducting gap in the quantum spin Hall edge state of monolayer WTe$_2$," *Nat. Phys.*, vol. 16, no. 5, Art. no. 5, May 2020, doi: 10.1038/s41567-020-0816-x.
[4]  Y. Shi *et al.*, "Imaging quantum spin Hall edges in monolayer WTe$_2$," *Science Advances*, vol. 5, no. 2, p. eaat8799, Feb. 2019, doi: 10.1126/sciadv.aat8799.
[5]  Z. Fei *et al.*, "Edge conduction in monolayer WTe$_2$," *Nature Physics*, vol. 13, no. 7, Art. no. 7, Jul. 2017, doi: 10.1038/nphys4091.
[6]  S. Wu *et al.*, "Observation of the quantum spin Hall effect up to 100 kelvin in a monolayer crystal," *Science*, vol. 359, no. 6371, pp. 76–79, Jan. 2018, doi: 10.1126/science.aan6003.
[7]  W. Zhao *et al.*, "Determination of the spin axis in quantum spin Hall insulator monolayer WTe$_2$," *arXiv:2010.09986 [cond-mat]*, Mar. 2021, Accessed: Mar. 18, 2021. [Online]. Available: http://arxiv.org/abs/2010.09986
[8]  S. Nandy and D. Pesin, "Low-energy effective theory and anomalous Hall effect in monolayer WTe$_2$," *SciPost Physics*, vol. 12, no. 4, p. 120, Apr. 2022, doi: 10.21468/SciPostPhys.12.4.120.
[9]  B. Sun *et al.*, "Evidence for equilibrium exciton condensation in monolayer WTe$_2$," *Nat. Phys.*, vol. 18, no. 1, Art. no. 1, Jan. 2022, doi: 10.1038/s41567-021-01427-5.
[10] E. Sajadi *et al.*, "Gate-induced superconductivity in a monolayer topological insulator," *Science*, vol. 362, no. 6417, pp. 922–925, Nov. 2018, doi: 10.1126/science.aar4426.
[11] V. Fatemi *et al.*, "Electrically tunable low-density superconductivity in a monolayer topological insulator," *Science*, vol. 362, no. 6417, pp. 926–929, Nov. 2018, doi: 10.1126/science.aar4642.

### 3.3 Two-dimensional Xenes

Harold J.W. Zandvliet and Pantelis Bampoulis

Physics of Interfaces and Nanomaterials, MESA+ Institute for Nanotechnology, University of Twente, P.O. Box 217, 7500AE Enschede, The Netherlands

**Status**

Xenes are two-dimensional (2D) monoelemental materials with a honeycomb graphene-like structure. 2D Xenes based on the group IVA elements Si (silicene), Ge (germanene), Sn (stanene), and Pb (plumbene) are isoelectric to graphene, but their honeycomb structure is buckled rather than flat, see Fig. 1a. Spin-orbit coupling (SOC) opens an energy gap at the Dirac (K, K') points. This band-inverted gap constitutes Xenes as 2D topological insulators (2DTI) and a platform to realize the quantum spin Hall effect (QSHE) [1]. The QSHE is characterized by an energy gap in the interior of the material and two counter-propagating, topologically protected helical edge states, Fig. 1a. Time-reversal symmetry and spin-momentum locking protects the edge states from backscattering, allowing for dissipationless electronic transport along the edges of the material. The realization of the QSHE at room temperature has significant practical implications since it can provide ideal conduction paths that are impervious to disorder.

Graphene has a very small SOC, necessitating extremely low temperatures to investigate the QSHE. In contrast, 2D Xenes formed with heavier elements have considerably wider topological gaps due to greater SOC (the SOC scales with the atomic number as $Z^4$). 2D Xenes have been epitaxially grown on a variety of substrates (see review [2] and references therein). Buckled and flat 2D Xene phases have been confirmed by scanning tunneling microscopy studies in conjunction with ab initio calculations [2-4]. Their electronic band structure has been also studied with angle-resolved photoemission spectroscopy and scanning tunneling spectroscopy. It's debated whether interactions with the substrate affect Xenes' topological nature. Strong interactions with metallic substrates are in general detrimental, but for silicene and stanene, they were found to be beneficial as they lead to unusual Dirac cones in silicene [5], and a large topological gap in stanene (three times larger than in free-standing stanene) [4]. Despite progress, major challenges remain in the development and characterization of 2D Xenes, which will be discussed in this roadmap.

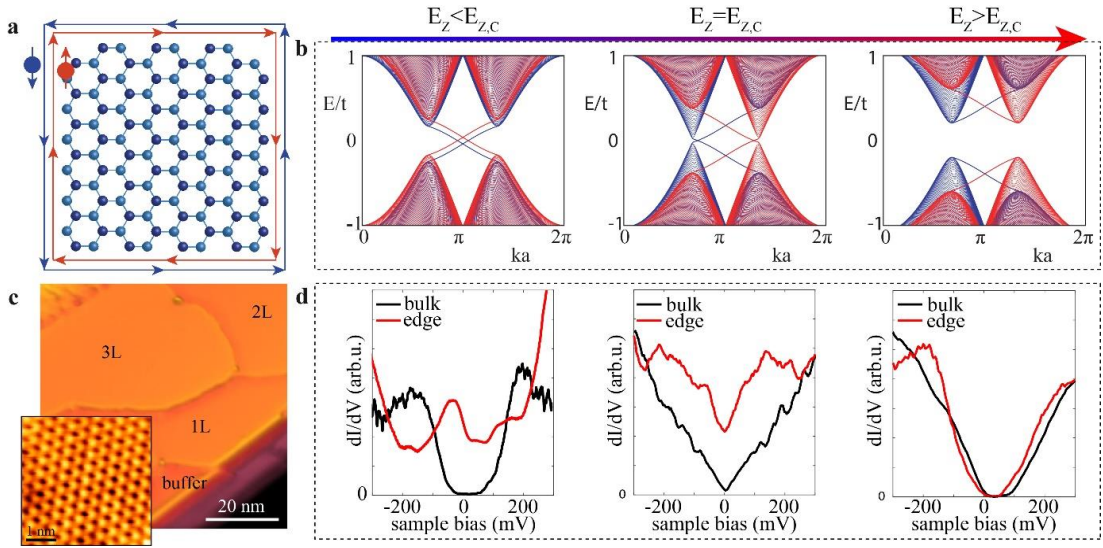

*Fig. 1. (a) Top view of the buckled honeycomb lattice of a 2D Xene. The different shades of blue indicate the two vertically displaced sublattices. The counter-propagating and spin-polarized edge states are depicted. (b) The QSH insulator's band structure under a perpendicular electric field, for the sub-critical field (Ez<Ez,c left), the critical field ((Ez=Ez,c middle) and the supercritical field ((Ez>Ez,c right), for which the material becomes a trivial insulator. Red/blue bands correspond to spin-up/down electrons. (c) STM image of few-layer germanene on Ge2Pt(101), the number of layers is indicated. Inset: high resolution STM image of the honeycomb lattice of the first decoupled germanene layer. (d) dI(V)/dV spectra recorded at the bulk (black) and edge (red) of germanene for electric a subcritical field (left), the critical field (middle) and a supercritical field (right), showing the transition from a QSH insulator to a topological semimetal and to a trivial insulator. The figure has been adapted from Ref. [8].*

**Current and Future Challenges**

2D Xenes is a rapidly evolving field. Because of their large gap and robust QSH states, they are not only a platform for fundamental research but are also promising for a wide range of applications. The buckling in 2D Xenes allows tuning of their quantum state of matter, e.g., by strain, functionalization, or an electric field [2]. An electric field-induced topological phase transition is of particular interest for the realization of a topological field effect transistor (TFET)[6-7]. In the 'ON' state of a TFET, current flows without energy losses along the protected edge channels. When a critical electric field is applied, the conducting channels vanish and the transistor is switched 'OFF'. Despite the frequent emergence of new paradigms, such as the TFET, since its very inception, the field has faced significant challenges:

(a) The growth of 2D Xenes is mainly based on UHV MBE processes on (mainly) metal substrates. This is far from ideal, the 2D layers are often small in size and interact strongly with the substrate. To make further scientific and technological advances, large-scale, high-quality single-crystal production methods are needed. For device development, techniques to isolate and safeguard the air-sensitive 2D Xenes are necessary.
(b) While robust edge states have been identified in germanene and stanene, the expected dissipationless charge transport at these edge states has not yet been experimentally verified.
(c) Just recently the possibility to change the topological phase using an external electric field was confirmed for germanene, see Figure 1(b)-(d) [8]. The application of an electric field allows to

    the change the quantum state of matter of germanene from a 2D topological insulator to a trivial band insulator. Scanning tunnelling spectroscopy as a function of an external electric field reveal that the band gap in the interior of germanene first closes and then reopens again. The reopening of the band gap in the interior of germanene goes hand in hand with the disappearance of the edge states, which provides compelling evidence for the topological nature of the edge states. These results [8] bring the realization of a TFET operating at room temperature within reach. Despite this advancement, the topological phase transition in germanene was very local. For device concepts, non-local approaches need to be taken, e.g. by using a gate voltage as in conventional transistors. Moreover, further research is required to reduce the required critical fields for the topological transitions.

(d) Identification of emerging application domains where 2D Xenes can play a crucial role and expanding the (topological) properties of 2D Xenes and their derivatives. Perhaps the most promising application of 2D Xenes is in TFETs. Helical edge states may be also useful for 1D ballistic interconnects and spintronics, but these ideas require further experimental investigation and effective control of topological edge states and spin.

**Advances in Science and Technology to Meet Challenges**

The scientific and technological underpinnings of 2D Xenes have not yet advanced to the point where they might be used in topological electronics. From a scientific standpoint, further research is required to optimize their growth. Growth optimization would improve their crystal quality, which could increase the prominence of QSHE at elevated temperatures and prevent doping. One could envision a menagerie of electronic properties by further engineering the quantum state of 2D Xenes. This could be done through surface functionalization, heterostructure combinations, proximity effects (by appropriate material combinations), and dimensionality reduction (e.g. nanoribbons) (see refs [2, 9] and references therein). Most current studies are limited to STM and ARPES but transport information is a necessity to first and foremost prove the dissipationless charge transport along the 2D Xenes' edge states. A four-point probe STM could be used to investigate charge transport at the edges of 2D Xenes, with two STM tips (electrodes) positioned along the 2D Xene edge and the other STM tip(s) used as probe(s) to intercept the conductive channels. These findings will offer insights into topological transport in 2D materials, with enormous scientific and technological ramifications.

New theoretical models must be developed to account for the complexity of 2D Xenes. This would provide relevant information to experimentalists, and predict the emergence of even more exotic states of matter. It will also provide a deeper understanding of topological phase transitions, the field-effect, the influence of SOC to the edge states as well as interactions between trivial and topological carriers. The latter is important, especially considering that perturbations of 2D Xenes by metals or semiconductors are unavoidable in view of transport experiments and device fabrication.

From a technological standpoint, the development of large band gap, decoupled 2D Xenes with high mobilities and low defect densities fabricated on a wafer scale will significantly advance the field. It will enable the development of proof-of-concept devices exploiting topological charge and spin transport. Another challenge for this emerging field is the development of device manufacturing techniques. Novel UHV-based processes utilizing prefabricated nanomasks, nanopatterning, and capping are as of now the sole viable option to combat the 2D Xene instability in ambient conditions. The initial steps in this direction have already been reported [10]. In the future, in order to integrate 2D Xenes into CMOS technologies, these approaches must ultimately become compatible with existing industry standards.

**Concluding Remarks**

The discoveries of the quantum spin Hall effect in germanene, large topological gaps in stanene and bismuthene, and Dirac cones in silicene have led to intense theoretical and experimental research on Xenes. With the possibility to tune the topological phase of 2D Xenes with external electric or magnetic fields, Xenes are great candidates for topological electronics. These fascinating features highlight the potential of 2D Xenes for enabling innovative applications in the field of materials science and technology. To fully realize their potential and enable novel and exciting applications, it is crucial to overcome upcoming challenges in growth, stability, scalability, and theoretical modelling. Despite these challenges, the future of 2D Xenes is promising, and we can expect significant advancements and innovations in the coming years.

**Acknowledgments**

H.J.W.Z. and P.B. acknowledge NWO for financial support (FOM 16PR3237, NWO OCENW.M20.232 and NWO VENI).

**References**

[1] C. L. Kane and E. J. Mele, "$Z_2$ Topological Order and the Quantum Spin Hall Effect," *Phys Rev Lett*, vol. 95, no. 14, p. 146802, Sep. 2005, doi: 10.1103/PhysRevLett.95.146802.

[2] A. Molle, J. Goldberger, M. Houssa, Y. Xu, S.-C. Zhang, and D. Akinwande, "Buckled two-dimensional Xene sheets," *Nat Mater*, vol. 16, no. 2, pp. 163–169, Feb. 2017, doi: 10.1038/nmat4802.

[3] P. Bampoulis, L. Zhang, A. Safaei, R. van Gastel, B. Poelsema, and H. J. W. Zandvliet, "Germanene termination of $Ge_2Pt$ crystals on Ge(110)," *J Phys Condens Matter*, vol. 26, no. 44, p. 442001, Nov. 2014, doi: 10.1088/0953-8984/26/44/442001.

[4] J. Deng *et al.*, "Epitaxial growth of ultraflat stanene with topological band inversion," *Nat Mater*, vol. 17, no. 12, pp. 1081–1086, Dec. 2018, doi: 10.1038/s41563-018-0203-5.

[5] Y. Feng *et al.*, "Direct evidence of interaction-induced Dirac cones in a monolayer silicene/Ag(111) system," *Proc Natl Acad Sci*, vol. 113, no. 51, pp. 14656–14661, Dec. 2016, doi: 10.1073/pnas.1613434114.

[6] N. D. Drummond, V. Zólyomi, and V. I. Fal'ko, "Electrically tunable band gap in silicene," *Phys Rev B*, vol. 85, no. 7, p. 075423, Feb. 2012, doi: 10.1103/PhysRevB.85.075423.

[7] M. Ezawa, "Monolayer Topological Insulators: Silicene, Germanene, and Stanene," *J Phys Soc Japan*, vol. 84, no. 12, p. 121003, Dec. 2015, doi: 10.7566/JPSJ.84.121003.

[8] P. Bampoulis *et al.*, "Quantum spin Hall states and topological phase transition in germanene," *Phys. Rev. Lett. 130, 196401 (2023), doi: 10.1103/PhysRevLett.130.196401*

[9] F. Bechstedt, P. Gori, and O. Pulci, "Beyond graphene: Clean, hydrogenated and halogenated silicene, germanene, stanene, and plumbene," *Prog Surf Sci*, vol. 96, no. 3, p. 100615, Aug. 2021, doi: 10.1016/j.progsurf.2021.100615.

[10]   L. Tao *et al.*, "Silicene field-effect transistors operating at room temperature," *Nat Nanotechnol*, vol. 10, no. 3, pp. 227–231, Mar. 2015, doi: 10.1038/nnano.2014.325.

## 3.4 Bismuthene – an atomic layer as large gap quantum spin Hall insulator

Ralph Claessen

*Physikalisches Institut and Würzburg-Dresden Cluster of Excellence ct.qmat, 97074 Würzburg, Universität Würzburg, Germany*

**Status**

As Kane and Mele have shown in their seminal work on graphene [1], atomic honeycomb lattices tend to form quantum spin Hall insulators (QSHI). However, with spin-orbit coupling (SOC) in graphene being too weak to open an appreciable band gap, honeycomb lattices composed of heavier group IV elements have come into focus as possible QSHI realizations (e.g., silicene or germanene – as discussed in section 3.3). A viable alternative are group V monolayers on wide-gap semiconductors. A case in point is Bi/SiC(0001), i.e., bismuthene, a honeycomb monolayer of Bi atoms covalently bonded to the SiC substrate [2]. It forms an atomic √3x√3 reconstruction with respect to the substrate lattice and exhibits a remarkably large band gap of 0.8 eV, the largest to date in any QSHI (see Fig. 1). The latter is attributed to "orbital filtering" in which the Bi $6p_z$ orbital is consumed by covalent bonding to the SiC substrate, leaving the in-plane Bi $6p_{x,y}$ orbitals to form the Dirac bands. Their double degeneracy allows for intra-atomic SOC (unlike graphene, whose Dirac band is derived from the single C $2p_z$ orbital), which explains the enormous gap size in addition to the high atomic number of Bi [3].

The key signature of a QSHI is the existence of one-dimensional (1D) metallic edge states in the bulk band gap. For bismuthene these have been confirmed by scanning tunnelling microscopy (STM) at monolayer terminations, e.g., near terrace steps of the substrate (Fig. 1c). They display an extreme spatial 1D confinement on atomic length scales, consistent with the large band gap [2]. Due to spin-momentum locking the topological edge states are protected against back-scattering off non-magnetic impurities, resulting in the absence of quasiparticle interference (QPI) in the STM data. Topological protection is lifted, however, at domain boundaries induced by the √3x√3 reconstruction (Fig.1b and e), which can be viewed as edge pairs coupled by electron tunnelling [4]. Because the relationship between spin orientation and momentum direction is reversed on opposite edges, back-scattering at defects and hence QPI get restored, leading to Fabry-Pérot like electron resonances between neighboring scatterers (Fig. 1e).

The 1D character of the topological edge states is also reflected in a prominent zero bias anomaly of their tunnelling spectra (Fig. 1c). Its specific energy and temperature dependence is characteristic of interacting 1D electrons, including universal scaling expected for a (helical) Tomonaga-Luttinger liquid (Fig. 1d). This demonstrates that the edge electrons – while topologically protected against single-particle backscattering – are still subject to two-particle scattering. The interaction strength extracted from these spectra (quantified by the Luttinger parameter K=0.42) places the bismuthene edge states in the intermediate to strong coupling regime, being a direct consequence of their narrow (atom-scale) confinement [5].

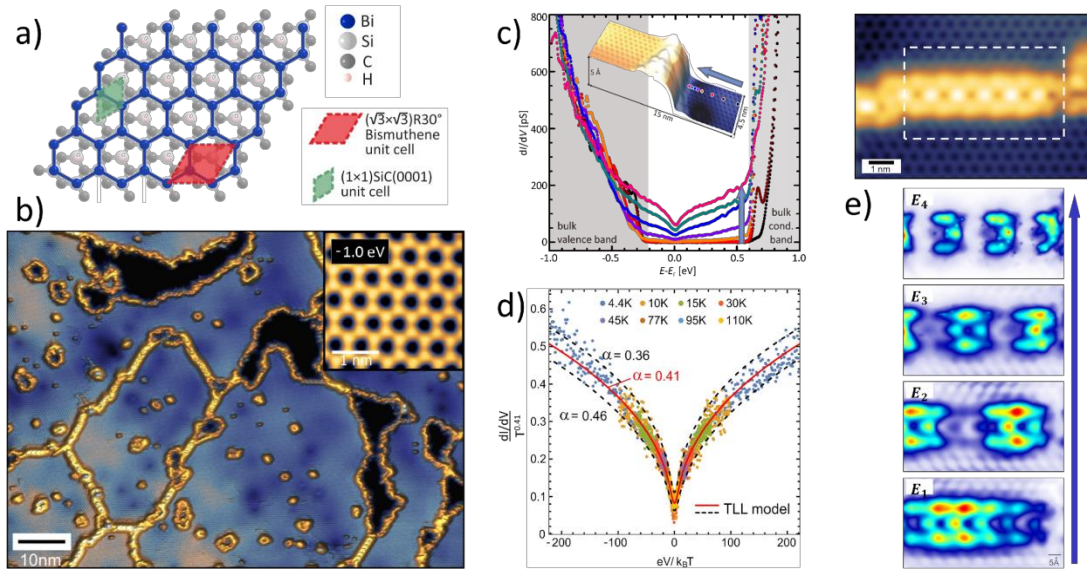

**Figure 1.** a) Schematic structure of bismuthene on SiC(0001) [2]. b) Large scale STM topography of bismuthene. The yellow lines are domain boundaries and free edges next to uncovered substrate surface (black). Inset: blow-up showing the atomic honeycomb arrangement. c) Local tunnelling spectra of the bulk band gap (black curve) and its gradual filling by metallic edge states when moving the tunnelling tip closer to a film edge at a substrate terrace step (see inset) [2]. d) Scaling of the zero bias anomaly seen in c) indicative of Tomonaga-Luttinger liquid behavior [5]. e) atomically resolved STM topography of a domain boundary section between two kinks (top panel) and spatial LDOS maps showing the formation of electronic Fabry-Pérot resonances at discrete energies (bottom panels).

**Current and Future Challenges**

The enormous band gap of bismuthene opens up new experimental opportunities that were not previously possible with 2D topological insulators. For example, it allows, in principle, edge transport experiments at room temperature without the interference of thermally activated bulk carriers, while transport studies of other 2D topological insulators with their much smaller gaps (a few 10 meV to 100 meV at most) have so far been limited to cryogenic temperatures. It would indeed represent a milestone to confirm the non-trivial topological character of bismuthene by a direct demonstration of the quantum spin Hall effect under ambient conditions. However, meaningful transport experiments are strongly hampered by bismuthene's domain structure (Fig.1b). Typical domain sizes are limited to well below 100 nm, implying that even micron-sized transport device structures are interspersed with a large number of domain boundaries, turning clean access to ballistic and non-local edge transport into a major challenge.

Another consequence of the large band gap is bismuthene's accessibility to optical experiments. In an initial application, a near-infrared pump pulse was used to excite electrons from the valence band into the conduction band, from which the electrons were photoemitted by a subsequent vacuum ultraviolet probe pulse [6]. This allowed complete band mapping, providing experimental proof of the indirect nature of the fundamental band gap. Moreover, the application of a time delay between the pump and probe pulses provides access to photocarrier lifetimes which are found to be an order of magnitude smaller than in trivial 2D semiconductors, attributed to enhanced deexcitation via the topological in-gap states at edges and domain boundaries [6]. Bismuthene's band structure also features a *direct* optical gap of approx. 1.4 eV at the K/K'-points of its hexagonal Brillouin zone, similar to that found in other 2D semiconductors and thus providing a platform for exciton formation. Indeed, the room-temperature optical response measured by laser-modulated photo-reflectivity reveals two

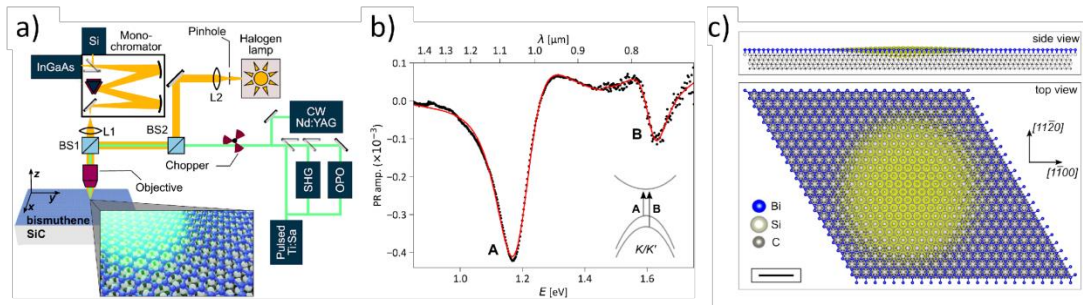

**Figure 2.** a) Experimental set-up for photo-modulated reflectivity measurements of bismuthene. b) Experimental spectrum (black points) and phenomenological line-shape fit (red curve) showing two resonances labelled A and B which are identified as excitons resulting from optical transitions at the K-points of the Brillouin zone (inset). c) Side and top view of the excitonic wave function as calculated by the Bethe-Salpeter equation. (from Ref. [7])

prominent resonances that could be unambiguously identified as excitation into bound electron-hole pairs with an excitonic binding energy of about 0.15 eV [7] (see Fig. 2). This first observation of excitons in a QSHI raises the question to what extent their properties are affected by the topological nature of the underlying band structure, potentially laying the foundation for novel "topological valleytronics". A recent theoretical study suggests that exciton formation in bismuthene may open new routes to create and manipulate qubit systems in 2D semiconductors [8].

**Advances in Science and Technology to Meet Challenges**

Now that our fundamental understanding of bismuthene has reached a mature stage, the focus is shifting to exploring its application potential in electronic and optoelectronic device structures. There are two main challenges here: Domain structure and chemical stability under ex situ conditions. Regarding the first challenge, previous attempts to optimize domain growth seem to have reached a limit, which still prevents mono-domain devices and thus uninterrupted ballistic edge transport. Alternative strategies for domain maximization rely on intentionally miscut SiC substrates for controlled terracing or the application of external strain but have yet to be implemented and evaluated. In fact, due to the √3x√3 symmetry of bismuthene's honeycomb lattice domain formation is in principle unavoidable. On the other hand, a denser Bi monolayer with a 1:1 atomic bonding to the topmost Si layer of the substrate (and hence single-domain layer) would result in a triangular lattice and thus an apparent loss of the Dirac semimetal character intrinsic to the honeycomb arrangement. Recently, however, it has been discovered that a triangular lattice can accommodate emergent honeycomb physics through appropriate orbital engineering, as recently demonstrated for indenene, a triangular In atom monolayer on SiC(0001). Here, the in-plane 5p orbitals of the In atoms form orbital angular momentum eigenstates leading to a topologically nontrivial band structure [9].

The second challenge concerns the chemical fragility of the atomic monolayers. Both bismuthene and indenene have van der Waals-like surfaces, but are still sensitive to rapid oxidation when exposed to ambient atmosphere. They hence require suitable protection when taken out of their ultra-high vacuum (UHV) birthplace, e.g., for ex-situ nanofabrication or actual device applications. At the same time any protective cap must leave the intrinsic electronic properties, especially the QSHI character, unaffected. This can be achieved by coating with van der Waals materials. For example, recent experiments have shown that a monolayer of graphene is a highly effective protection of the QSHI indenene against oxidation and even exposure to liquid water [10]. For transport device applications and optical experiments one would, however, prefer a large gap insulator as protective layer, such as hexagonal boron nitride (hBN). The technology for controlled hBN capping is well developed and in standard use for other 2D semiconductors. Its extension to UHV-born atomic monolayer systems like bismuthene is, in principle, straightforward and currently under development.

**Concluding Remarks**

Bismuthene has been the first atomic monolayer system experimentally demonstrated to be a QSHI. Its record band gap sets it apart from all other known QSHI materials and offers unprecedented experimental opportunities and ideas for new applications, especially with respect to its optical response. In contrast, bismuthene's inherent propensity to form structural domains appears to be a major obstacle to exploiting its topological edge transport. On the other hand, the discovery of indenene, a triangular atomic layer with emergent honeycomb physics, opens the QSHI realm to a much wider class of atomic monolayer systems yet to be explored. Together with the technological developments for protective device integration of otherwise fragile atomic layers, bismuthene and indenene can be considered as representatives of a third generation of 2D quantum materials with promising application potential – after graphene and single-layer transition metal dichalcogenides.

**Acknowledgements**
I wish to thank all coauthors of Refs. [2]-[7], [9], and [10] for the fruitful collaboration, and especially R. Stühler for his central contributions. Key funding from the Deutsche Forschungsgemeinschaft (DFG, German Research Foundation) under Germany's Excellence Strategy through the Würzburg-Dresden Cluster of Excellence EXC 2147 on Complexity and Topology in Quantum Matter ct.qmat (project ID 390858490) as well as through the Collaborative Research Center SFB 1170 ToCoTronics (project ID 258499086) is gratefully acknowledged.

**References**

[1] Kane C L and Mele E J 2005, Quantum spin Hall effect in graphene *Phys. Rev. Lett.* **95** 226801
[2] Reis F, Li G, Dudy L, Bauernfeind M, Glass S, Hanke W, Thomale R, Schäfer J, and Claessen R 2017 Bismuthene on a SiC substrate: A candidate for a high-temperature quantum spin Hall material *Science* **357**, 287
[3] Li G, Hanke W, Hankiewicz E M, Schäfer J, Claessen R, Wu C-J, and Thomale R (2018) Theoretical paradigm for the quantum spin Hall effect at high temperatures *Phys. Rev. B* **98** 165146
[4] Stühler R, Kowalewski A, Reis F, Jungblut D, Dominguez F, Scharf B, Li G, Schäfer J, Hankiewicz E M, and Claessen R (2022) Effective lifting of the topological protection of quantum spin Hall edge states by edge coupling *Nat. Commun.* **13** 3480
[5] Stühler R, Reis F, Müller T, Helbig T, Schwemmer T, Thomale R, Schäfer J, and Claessen R (2020) Tomonaga-Luttinger liquid in the edge channels of a quantum spin Hall insulator *Nat. Phys.* **16** 47
[6] Maklar J, Stühler R, Dendzik M, Pincelli T, Dong S, Beaulieu S, Neef A, Li G, Wolf M, Ernstorfer R, Claessen R, and Rettig L (2022) Ultrafast momentum-resolved hot electron dynamics in the two-dimensional topological insulator bismuthene *Nano Lett.* **22** 5420
[7] Syperek M, Stühler R, Consiglio A, Holewa P, Wyborski P, Dusanowski Ł, Reis F, Höfling S, Thomale R, Hanke W, Claessen R, Di Sante D, and Schneider C (2022) Observation of room temperature excitons in an atomically thin topological insulator *Nat. Commun.* **13** 6313
[8] Ruan J, Li Z, Ong C S, and Louie S G (2022) Two-dimensional single-valley exciton qubit and optical spin magnetization generation *arXiv:2211.03334*

[9]   Bauernfeind M, Erhardt J, Eck P, Thakur P K, Gabel J, Lee T-L, Schäfer J, Moser S, Di Sante D, Claessen R, and Sangiovanni G (2021) Design and realization of topological Dirac fermions on a triangular lattice *Nat. Commun.* **12**, 5396

[10]  Schmitt C, Erhardt J, Eck P, Schmitt M, Lee K, Wagner T, Keßler P, Kamp M, Kim T, Cacho C, Cephise, Lee T-L, Sangiovanni G, Moser S, and Claessen R (2023) Stabilizing an atomically thin quantum spin Hall insulator at ambient conditions: graphene-intercalation of indenene *arXiv:2305.07807*

## 3.5 Quantum anomalous Hall insulators

*Fabian R. Menges[1], Johannes Gooth[1], Claudia Felser[1,2] and Chandra Shekhar[1]*
[1]Max Planck Institute for Chemical Physics of Solids, 01187 Dresden, Germany
[2] Würzburg-Dresden Cluster of Excellence ct.qmat, Universität Würzburg, 97074 Würzburg, Germany

**Status**
The quantum Hall effect (QHE) has revolutionised our understanding of electronic states in condensed matter physics. But how can quantum Hall states with their characteristic dissipationless chiral transport nature be realized in the absence of high magnetic fields and at ambient temperatures? This is a key question in today's topological materials science, and outlines fascinating directions for future research on quantum anomalous Hall insulators.

First observed in a two-dimensional (2D) electron gas subjected to strong magnetic fields and temperatures close to absolute zero, the QHE exhibits a quantised Hall conductance on the order of $e^2/h$, accompanied by vanishing longitudinal resistance [1]. The topological nature of the effect was elucidated by Laughlin through the concept of Chern numbers, establishing a link between the exact quantisation and the topological invariance of the energy bands. Subsequently, quasi-quantised Hall conductance has been observed beyond the original 2D electron gas, including the 3D Dirac semimetal $Cd_3As_2$ [2] and the quasi-3D semimetal $ZrTe_5$ [3], using different underlying principles such as the Fermi arcs and Fermi vectors. In graphene, the QHE has been observed up to room-temperature in high-magnetic fields > 20T [4].

The first theoretical concepts to realise QH states without the need for high magnetic fields date back to the late 1980s. Haldane proposed the prediction of the quantum anomalous Hall effect (QAHE) [5]. While it was not immediately clear how a QAHE could be realised in practice, its classical counterpart, the anomalous Hall effect (AHE) was well known to be a characteristic feature of many ferromagnetic materials, where the moving electrons gain additional velocity due to spontaneous magnetisation, which acts as a fictitious field. However, it was not until the discovery of the QHE, before these anomalous velocities were incorporated into the concept of Berry curvature.

Progress in this field advanced with the discovery of topological insulators (TIs). TIs can host QAH states when doped with magnetic elements such as Cr or V in (Bi, Sb, Te) [6, 7]. However, QAH states in these systems can currently only be realised at mK temperatures. Alternative strategies are, therefore, needed to extend the range of materials capable of hosting QAH states at higher temperatures.

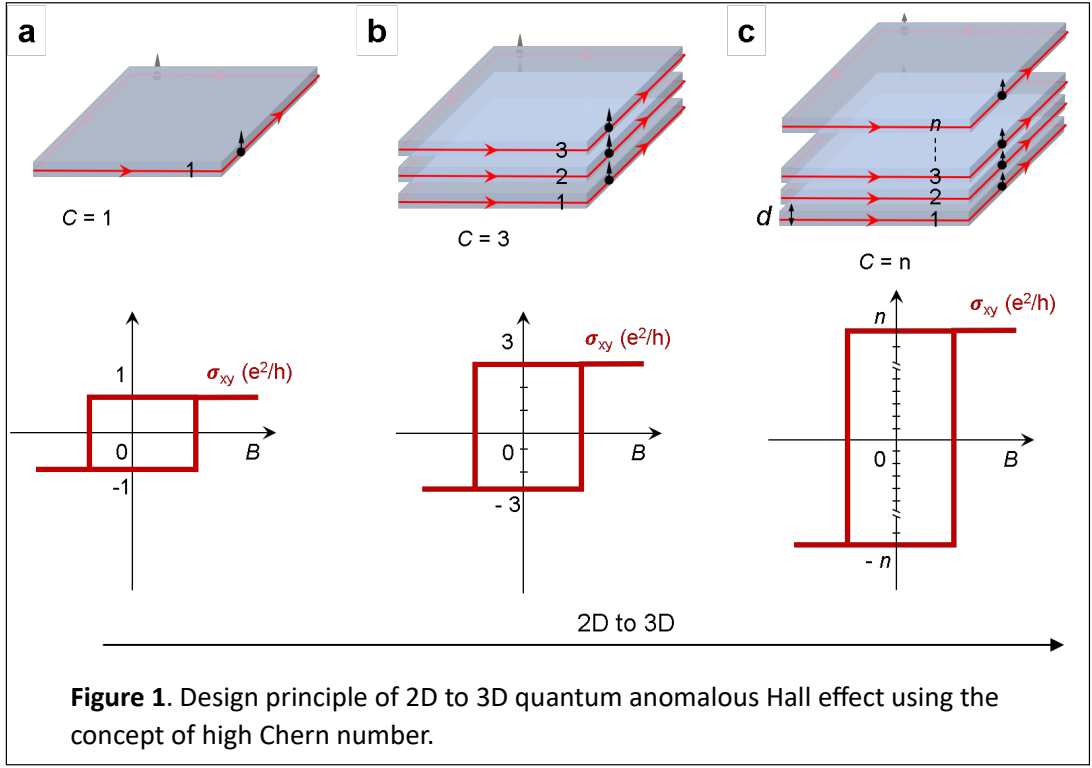

**Figure 1**. Design principle of 2D to 3D quantum anomalous Hall effect using the concept of high Chern number.

**Current and Future Challenges**

For the prototype examples of 2D QAH insulators, such as $Cr_{0.15}(Bi_{0.1}Sb_{0.9})_{1.85}Te_3$ and $V_{0.11}(Bi_{0.29}Sb_{0.71})_{1.89}Te_3$, the underlying mechanism of perfect quantisation can be understood by considering a single chiral conducting channel with a Chern number $C$ equal to 1 [6, 7]. This insight has been further validated and generalised in recent studies reporting the design of Hall insulators with higher-$C$ numbers based on multilayers consisting of Cr-doped $(Bi,Sb)_2Te_3$ as the Hall conducting layer and $(Bi,Sb)_2Te_3$ as the insulating spacer layer. The Chern number in such multilayer configurations can be tuned by adjusting the thickness or the magnetic Cr doping concentration [8], which requires complex materials engineering. While these studies provided proof-of-concept for the correlation between multilayer stacking and enhancing $C$, the observation of the QAHE in these systems is currently limited to mK temperatures. More recently, QAH states have been discovered in the intrinsic magnetic TI $MnBi_2Te_4$, even at higher temperatures reaching up to 1.5K. Inspired by the stacking of magnetic multilayers of TIs and the tunability of $C$ in $MnBi_2Te_4$, an efficient approach has been developed to realise high-$C$ states in MnBi2Te4/hBN multilayer heterostructures. Stacking of $n$ layers of $MnBi_2Te_4$ films with $C = 1$, interspersed with hBN monolayers, results in a high-$C$ multilayer structure characterised by $n$ chiral edge modes [9]. In this configuration, the total conductivity can be defined as $\sigma = C\frac{e^2}{h}\frac{1}{d}$, where $d$ is the distance between two stacked layers. The individual layers of $MnBi_2Te_4$ are effectively separated and behave as Hall insulators with only weak interfere with neighbouring layers. These concepts can probably be extended to certain 3D compounds where the ferromagnetic layers are well defined and separated by nonmagnetic spacer-layers [10]. Two prototype ferromagnetic compounds where this approach may be applicable in $Co_3Sn_2S_2$ and MnAlGe. Here, the spins of the Co and Mn atoms are ferromagnetically aligned out of plane within each layer, and these layers are separated by the nonmagnetic spacer layers.

**Advances in Science and Technology to Meet Challenges**

A promising path forward seems to be the study of topological magnetic compounds that exhibit a large AHE up to the magnetic transition. For example, $Co_3Sn_2S_2$ and MnAlGe exhibit AHC values of 1100 $\Omega^{-1}cm^{-1}$ and 670 $\Omega^{-1}cm^{-1}$, respectively, with corresponding transition temperatures ($T_C$) of 180 K and 500 K [11,12]. The AHC of $Co_3Sn_2S_2$ exhibits perfect square loop behaviour when a magnetic field is applied perpendicular to the magnetic layers. The distance between the two magnetic layers $d$, is 4.392 Å ($d = \frac{1}{3}c$, where $c$ is the lattice parameter of the hexagonal unit cell) in $Co_3Sn_2S_2$, while it is 5.93 Å ($d = c$) in MnAlGe. To determine the conductance per layer, the AHC is multiplied by $d$ in the units of $e^2/h$. The scaled AHC values closely approach the quantum conductance value of $e^2/h$ (= 3.874 x $10^{-5}$ $\Omega^{-1}$). Such a quantisation method is further validated by the discovery of the chiral edge states in $Co_3Sn_2S_2$ [13]. Remarkably, all the magnetic compounds that follow this $1/d$ scaling exhibit topologically non-trivial properties. This phenomenon can be understood by the stacking of 2D QAHE along the layer direction, considering that the overall topological phase is determined by the tunneling effect between neighboring 2D QAH states. In materials such as $Co_3Sn_2S_2$ and MnAlGe, the origin of the topological state is in the magnetic layers, where band inversion arises from intrinsic magnetisation and spin-orbital coupling. In certain cases, additional crystal symmetry, such as mirror reflection, can also lead to the formation of nodal line band structures, which exist in $Fe_3GeTe_2$. Interestingly, the individual magnetic layers within the bulk form of these compounds appears to fulfil the requirement for achieving the QAHE. It can be expected that thin films or monolayers of these compounds hold great promise for achieving QAHE at much higher temperatures than previously observed. However, it should be noted that the preparation of such films and monolayers is not only a complex process, but can also introduce changes in the physical properties of the compounds, including defects and alterations in the magnetic transition temperatures. These factors need to be carefully considered and further investigated to ensure the effectiveness of the proposed approach.

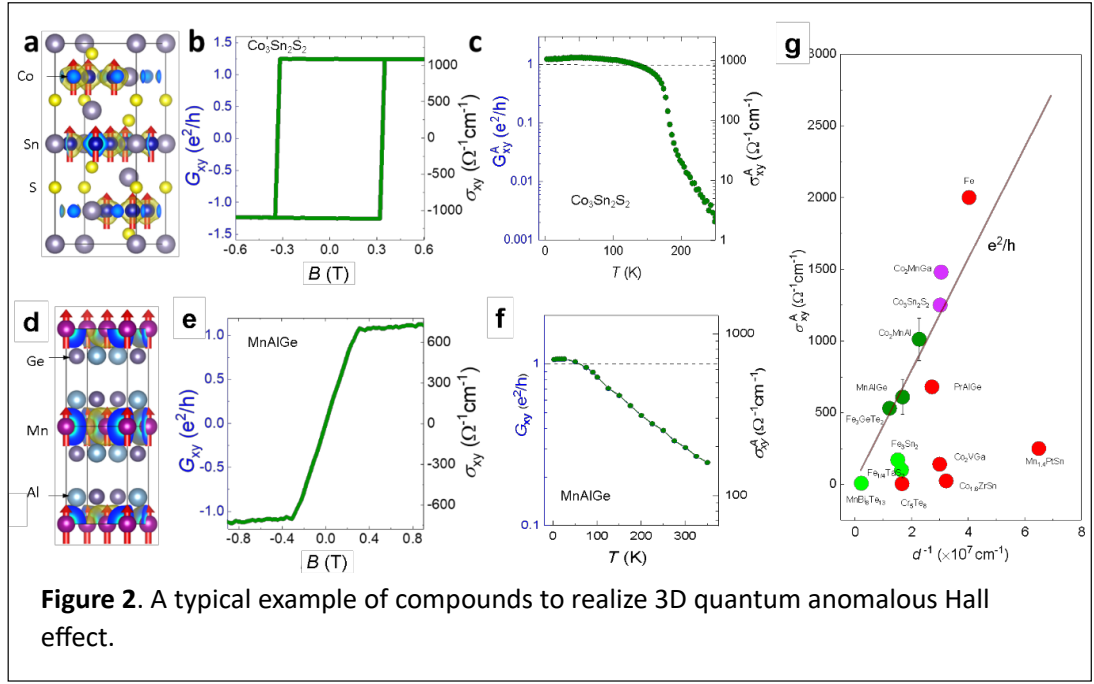

**Figure 2**. A typical example of compounds to realize 3D quantum anomalous Hall effect.

**Concluding Remarks**

By manipulating the Chern number in QAH insulators, the channel degree of freedom of the chiral edge states can be controlled, thereby creating opportunities to leverage topologically protected and dissipationless chiral edge states for quantum information storage and transfer. To date, the QAH insulator has only been observed either in thin films or in stacked multilayers at very low temperatures. One possibility is the high temperature topological layered magnets since their thickness scaling with $1/d$ coincides with the quantum conductance $e^2/h$. However, more in-depth experimental and theoretical research is needed to search for new materials and high-temperature magnetic topological insulators are one of the most promising families of materials.

**Acknowledgements**

This work was financially supported by the Deutsche Forschungsgemeinschaft (DFG) under SFB1143 (project no. 247310070) and the Würzburg-Dresden Cluster of Excellence on Complexity and Topology in Quantum Matter—ct.qmat (EXC 2147, project no. 390858490).

**References**

1. Klitzing KV, Dorda G, Pepper M 1980 New method for high-accuracy determination of the fine-structure constant based on quantized Hall resistance *Phys. Rev. Lett.* **45**, 494-497 (1980).
2. Zhang C, Zhang Y, Yuan X, Lu S, Zhang J, Narayan A, Liu Y, Zhang H, Ni Z, Liu R, Choi E S. 2019 Quantum Hall effect based on Weyl orbits in $Cd_3As_2$ *Nature* **565** 331–336.
3. Tang F, Ren Y, Wang P, Zhong R, Schneeloch J, Yang SA, Yang K, Lee PA, Gu G, Qiao Z, Zhang L 2019 Three-dimensional quantum Hall effect and metal–insulator transition in $ZrTe_5$ *Nature* **569** 537–541.
4. Novoselov KS, Jiang Z, Zhang Y, Morozov SV, Stormer HL, Zeitler U, Maan JC, Boebinger GS, Kim P, Geim AK 2007 Room-Temperature Quantum Hall Effect in Graphene *Science* **315** 1379.
5. Haldane F D 1988 Model for a quantum Hall effect without Landau levels: condensed-matter realization of the "parity anomaly" *Phys. Rev. Lett.* **61**, 2015-2018.
6. Chang C Z, Zhang J, Feng X, Shen J, Zhang Z, Guo M, Li K, Ou Y, Wei P, Wang L L, Ji Z Q 2013 Experimental observation of the quantum anomalous Hall effect in a magnetic topological insulator *Science* **340** 167–170.
7. Chang C Z, Zhao W, Kim DY, Zhang H, Assaf B A, Heiman D, Zhang S C, Liu C, Chan M H, Moodera J S 2015 High-precision realization of robust quantum anomalous Hall state in a hard ferromagnetic topological insulator *Nat. Mater.* **14** 473.
8. Zhao Y F, Zhang R, Mei R, Zhou L J, Yi H, Zhang Y Q, Yu J, Xiao R, Wang K, Samarth N, Chan M H 2020 Tuning the Chern number in quantum anomalous Hall insulators *Nature* **588** 419–423.
9. Bosnar M, Vyazovskaya A Y, Petrov E K, Chulkov E V, Otrokov M M 2023 High Chern number van der Waals magnetic topological multilayers MnBi2Te4/hBN *npj 2D Mater. Appl.* **7** 33.
10. Jin Y J, Wang R, Xia B W, Zheng B B, Xu H 2018 Three-dimensional quantum anomalous Hall effect in ferromagnetic insulators *Phys. Rev. B* **98** 081101.
11. Liu E, Sun Y, Kumar N, Muechler L, Sun A, Jiao L, Yang S Y, Liu D, Liang A, Xu Q, Kroder J *et al.* 2018 Giant anomalous Hall effect in a ferromagnetic kagome-lattice semimetal. *Nat. Phys.* **14** 1125-1131.
12. Guin SN, Xu Q, Kumar N, Kung H H, Dufresne S, Le C, Vir P, Michiardi M, Pedersen T, Gorovikov S, Zhdanovich S *et al.* 2021 2D-Berry-curvature-driven large anomalous Hall effect in layered Topological nodal-line MnAlGe *Adv. Mater.* **33** 200630.

13. Howard S, Jiao L, Wang Z, Morali N, Batabyal R, Nag P K, Avraham N, Beidenkopf H, Vir P, Liu E, Shekhar C *et al.* 2021 Evidence for one-dimensional chiral edge states in a magnetic Weyl semimetal $Co_3Sn_2S_2$ *Nat. Commun.* **12** 4269.

## 4.1 Angle-resolved photoemission spectroscopy (ARPES)

Anton Tadich (Australian Synchrotron), Mengting Zhao and Mark T. Edmonds (Monash University)

**Status**

Angle-resolved photoemission spectroscopy (ARPES) is one of the premier probes of electronic structure. By simultaneously measuring the photoelectron's kinetic energy and emission angle relative to crystallographic axes, it allows direct measurement of a material's electronic bandstructure – the binding energy as a function of the two-dimensional surface wavevector. Importantly, because ARPES measures the single particle, charged excitation spectrum, this allows the energy and momentum-resolved complex self-energy to be extracted. This spectrum encodes not only the bandstructure, but also the renormalization of states represented by the complex self-energy function, arising from spin, charge, and vibrational fluctuations. This has enabled ARPES to be central to the discovery of new topological systems such as directly observing the helical surface states of a 3D topological insulator [1], as well as observing 3D Dirac and Weyl cones, and the Fermi arc surface states of Weyl semimetals [2]. Furthermore, as most ARPES measurements typically only probe the first few atomic layers of a solid (due to the small inelastic mean free path of photoelectrons) it makes the technique ideally suited to the characterisation of 2D materials including two-dimensional topological insulators (2D TI).

In the field of 2D TI materials, ARPES has been instrumental in revealing details of the band structure of numerous systems including members of the honeycomb lattice family (e.g bismuthene [3]), the 1T' structural phases of transition metal dichalcogenides such as $WTe_2$ [4] and ultra-thin layers of $Na_3Bi$ [5]. The direct measurement of the momentum resolved electronic bandstructure below the Fermi level ($E_F$) allows fundamental material properties such as electron/hole doping, bandgap magnitude, carrier velocity and effective mass to be determined. Furthermore, ARPES allows observation of quantum phase transitions between trivial and non-trivial topological states (Fig. 1) [5], as well as band renormalisation and hybridisation effects and finally measurement of the orbital character of the 2D TI bulk band states. The latter property is a particularly powerful capability, allowing one to directly confirm topological band inversion.

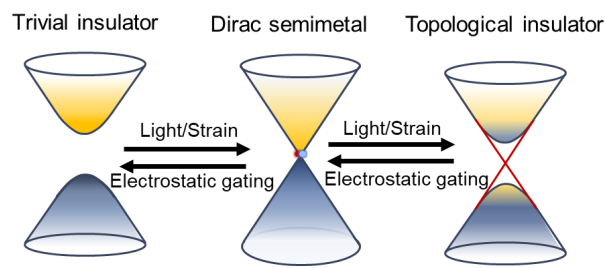

Figure 1. A schematic highlighting topological phase transitions between topologically trivial insulator and 2D topological insulator phase where the colour scale corresponds to the orbital character of the bands. These transitions can be induced via electric fields (gating or dosing), light and strain and can be utilised in an ARPES experiment to allow for direct spectroscopic visualisation of the topological phase transition.

Despite the detailed insights into electronic structure that ARPES can provide there are still several technical challenges that arise when applied to 2D TI systems. These challenges and the recent technological advancements in ARPES techniques that seek to overcome these challenges are outlined in the remaining sections.

**Current and Future Challenges**

There are currently three key challenges when using ARPES to characterise 2D TIs. The first is the preparation of atomically thin, clean and ordered crystals. Whilst molecular beam epitaxy (MBE) can grow highly ordered films with well-defined thickness, lattice mismatch with the substrate can lead to tensile or compressive strain resulting in detrimental changes to the bandstructure [3]. An alternative method is exfoliation of ultra-thin flakes from suitable bulk crystals [6]. However, the small size of these flakes and non-uniformity in their thickness present challenges when coupled with the relatively large spot size of conventional ARPES sources. Finally, without *in-situ* ARPES capability following sample preparation, it can be difficult to maintain the pristine sample quality needed for ARPES measurements when transporting to a synchrotron or laboratory facility.

The second challenge is observation of the 1D helical edge state itself. Compared to the topological surface states of 3D TIs which are present over the entire surface, for a 2D TI the topologically non-trivial 1D edge state is highly localized (within 1-2 nm) to the sample perimeter [3,4,5]. Whilst, the 1D edge state has yet to be observed with ARPES when employing nano/micro-ARPES on small exfoliated flakes or MBE films with a large number of small islands [4,6] the technique has previously been used to probe 1D metallic chains [7]. Therefore, development of a growth technique where an ordered TI is grown epitaxially on a vicinal substrate may enable ARPES to resolve the 1D edge state.

The final challenge is that ARPES is only able to measure occupied electronic states. For many 2D TI systems such as bismuthene [3] and $Na_3Bi$ [5], $E_F$ is found within the bulk band gap. To measure the size of the gap, electronic doping is required to observe the conduction band edge. In an ARPES experiment this is typically achieved via surface transfer doping using alkali metal adatoms. Aside from possible unwanted chemical reactions with the 2D TI, the process is fairly uncontrolled; the dopant atoms can cause surface disorder through clustering, compromising the ARPES data quality. Of more concern is that along with the rigid shift of the bands, changes to the dielectric screening environment can occur, leading to changes in the bandgap and the topological features.

**Advances in Science and Technology to Meet Challenges**
In recent years, major advances in ARPES instrumentation and sample environments are helping to address these challenges. Firstly, advanced sample encapsulation strategies (e.g. graphene overlayers) allow for transport of pristine exfoliated flakes [6], whilst compact UHV vacuum "suitcases" enable transport of MBE grown samples from home laboratories to ARPES facilities without exposure to air.

Technological progress in ARPES detectors and excitation sources (both synchrotron and laboratory) now allow experiments to be conducted with extremely high energy (few-meV) and momentum space ($\approx 0.01$ Å$^{-1}$) resolution. Spectrometers featuring high-efficiency spin-detection facilitate measurement of the complete 3D spin vector with high energy and angular resolution. The increasing brilliance of 3rd generation synchrotron sources in conjunction with advanced zone plate and capillary optics have resulted in powerful micro and nano-ARPES instruments capable of sub-micron spot sizes allowing spatially-resolved ARPES of micron-scale flakes. Whilst, laboratory ARPES systems employing intense, high harmonic generation pulsed laser sources have emerged allowing fs-scale time-resolved ARPES (tr-ARPES) measurements on 2D TIs such as 1T'-$WTe_2$ [8], allowing access to the unoccupied band states without the need for chemical doping. These advancements make it possible (at least in principle) to probe energy, momentum, spin, time and spatial resolution in a single photoemission spectroscopy setup as depicted schematically in Figure 2.

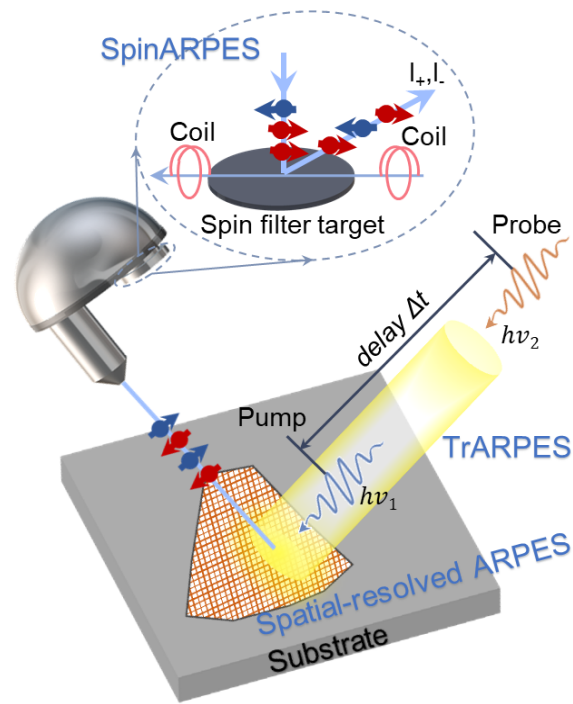

Figure 2. Schematic representation of the ability of ARPES measurements to probe not only energy and momentum resolved electronic states but also (1) Spin-resolved ARPES through the addition of a spin-detector, (2) Time-resolved ARPES with pulsed light sources by combing the pump-probe technique and (3) Spatially-resolved ARPES using micron-sized spot sizes.

Developments in micro/nano ARPES technology, as well as improvements in 2D heterostructure device fabrication, have also allowed in recent years electrostatic gating effects to be visualised with ARPES [9]. By employing a backgate and ultra-thin dielectrics in close proximity to the sample, large electric fields can be established at moderate gate voltages, rigidly shifting the bandstructure in a much more controllable and, importantly, reversible manner compared to alkali metal doping. This important scientific advancement means that electronic structure measurements can be carried out on actual current carrying electronic devices comprised of 2D TIs. Thus, providing an opportunity to measure spectroscopically controlled topological phase transitions via electrostatic gating.

Finally, as a solution to control and investigate the role of residual strain in ARPES measurements on epitaxially-grown 2D TIs, a number of setups have been developed which allow a reversible uniaxial strain to be applied to the sample [10]. This offers the ability to control the topological phase through a tailored strain-ARPES experiment. Whilst efforts until now have mostly focussed on phase transition behaviours in oxide materials, it should be straightforward to apply to the 2D TI family.

**Concluding Remarks**

The continued rapid advancement in ARPES detectors and light sources, as well new methods in sample preparation means many of the challenges highlighted above for 2D TI research using ARPES are now within reach. Yet, observation of the 1D edge state using ARPES remains a major and enduring challenge. Yet, there is hope this challenge can be overcome, as ARPES photoemission data has been successfully obtained from a genuine 1D metal grown on an insulating substrate [7]. If the challenge of measuring the 1D edge state with ARPES is overcome, it would represent a game-changing development for the field as it would enable a smoking-gun experiment to show spin-momentum coupling on the edge state. Further, it could also be used to study helical Luttinger liquid behaviour.

However, even if the 1D edge state remains elusive in an ARPES measurement, the expanding capabilities of modern ARPES instruments as well as techniques to tune electric field, strain and doping, still make ARPES a critical tool in discovering and understanding the electronic properties of new 2D TI candidates.

**Acknowledgements**

*M. T. E acknowledges funding from the Australian Research Council Future Fellowship program (FT220100290). M. Z. acknowledges funding from the ANSTO Postgraduate Fellowship. M. T. E., M. Z. and A. T. acknowledge support from the ARC Centre for Excellence Future Low Energy Electronic Technologies (FLEET) (CE1700100039).*

**References**

[1] D. Hsieh, Y. Xia, D. Qian, L. Wray, J. H. Dil, F. Meier, J. Osterwalder, L. Patthey, J. G. Checkelsky, N. P. Ong, A. V. Fedorov, H. Lin, A. Bansil, D. Grauer, Y. S. Hor, R. J. Cava, M. Z. Hasan, " A tunable topological insulator in the spin helical Dirac transport regime" *Nature* **460**, 1101-1105 (2009)

[2] S-Y. Xu, I. Belopolski, N. Alidoust, M. Neupane, G. Bian, C. Zhang, et al. "Discovery of a Weyl fermion semimetal and topological Fermi arcs" *Science* **349**, 613-617 (2015)

F. Reis, G. Li, L. Dudy, M. Bauernfeind, S. Glass, W. Hanke, *et. al.* "Bismuthene on a SiC substrate: A candidate for a high-temperature quantum spin Hall material." *Science* **357**, 6348 (2017) [4] S. Tang, C. Zhang, D. Wong, Z. Pedramrazi, H-Z Tsai, C. Jia, *et al*. "Quantum spin Hall state in monolayer 1T'-WTe$_2$" *Nature Phys* **13**, 683–687 (2017)

[5] J.L. Collins, A. Tadich, W. Wu, L.C. Gomes, J.N.B. Rodrigues, C. Liu, *et al*. "Electric-field-tuned topological phase transition in ultrathin Na$_3$Bi". *Nature* **564**, 390–394 (2018)

[6] I. Cucci, I. Gutiérrez-Lezama, E. Cappelli, S. M. Walker, F. Y. Bruno, G. Tenasini, *et. al.* "Microfocus Laser–Angle-Resolved Photoemission on Encapsulated Mono-, Bi-, and Few-Layer 1T'-WTe$_2$" *Nano Lett.* **19**, 1, 554–560 (2019)

[7] P. Segovia, D. Purdie, M. Hengsberger, Y. Baer, "Observation of spin and charge collective modes in one-dimensional metallic chains*" Nature* **402**, 504-507 (1999)

[8] E.J. Sie, T. Rohwer, C. Lee, N. Gedik. "Time-resolved XUV ARPES with tunable 24–33 eV laser pulses at 30 meV resolution" *Nat Commun* **10**, 3535 (2019)

[9] P.V. Nguyen, N.C. Teutsch, N.P. Wilson, J. Kahn, X. Xia, A. J. Graham, *et al.* "Visualizing electrostatic gating effects in two-dimensional heterostructures" *Nature* **572**, 220–223 (2019)

[10] S. Riccò, M. Kim, A. Tamai, S. McKeown Walker, F. Y. Bruno, I. Cucchi, *et al.* "In situ strain tuning of the metal-insulator-transition of Ca$_2$RuO$_4$ in angle-resolved photoemission experiments". *Nat Commun* **9**, 4535 (2018)

## 4.2 Scanning Tunneling Microscopy and Spectroscopy (STM/S)

Junxiang Jia[1], Bent Weber[1]
[1]Nanyang Technological University

**Status**

Scanning tunneling microscopy and spectroscopy (STM/STS) with its unrivalled spatial and spectral resolution can provide unprecedented insight into the local electronic structure of 2D topological insulators (TI) [1]. Demonstrations including precise measurements and even the tunability of bulk energy gaps [2–4], and visualizations of the definition, extent, and continuity of atomically confined topological edge states [3, 5–7]. More so, STM/STS has been shown to provide information beyond the single-particle picture, including the characterization of electronic correlations [6, 7] and superconductivity [8, 9].

Following the first scanning probe based confirmation of a QSH insulating 2D bulk and metallic boundary states in the low-energy electronic structure of WTe$_2$ by Tang et al. [5] (Figure 1A and 1B), similar signatures of bulk-boundary correspondence have been reported also for other QSH material candidates, such as the Xenes (germanene [4], buckled antimonene [10, 11], plumbene [12], bismuthene [13]), as well as atomically-thin layers of the 3D Dirac/Weyl semimetals Na$_3$Bi [2] and the transition metal dichalcogenides (TMDCs) [5]. A detailed summary of material systems and their 2D bulk gaps can be found in Ref. [1] and in other sections of this roadmap.

Measurement of the local density of states (LDOS) by STM/STS has made it possible not only to visualize topological edge state signatures (Figure 1C and 1D) [3, 6], but to infer detailed information on the presence and strength of electronic correlations.  Within the strictly 1D helical edge states of quantum spin Hall (QSH) insulators, a breakdown of screening [14] leads to the formation of a helical Tomonaga-Luttinger liquid (TLL). Its signature is a characteristic pseudogap-like suppression of the low-energy tunneling DOS at the Fermi energy. Recently confirmed by temperature-dependent STM/STS for the first time in the atomically-thin 2D TI bismuthene [6] (Figure 1E), this pseudogap scales "universally" as a power-law in both temperature and bias voltage, allowing to extract the Luttinger parameter $K$ which quantifies the correlation strength. In WTe$_2$ [7], it was subsequently shown that the strength of the many-body Coulomb interactions within the TLL can further be controlled via the edge's dielectric environment and the Fermi velocity [15] (Figure 1F), providing additional evidence of the Luttinger picture.

Particular to atomically-thin 2D TIs, their highly confined edge states can harbor exceptionally strong electronic correlations, reported as $K \simeq 0.42$ in bismuthene [6] and $K \simeq 0.2 - 0.4$ in WTe$_2$ [7] ($K = 1$ indicates a non-interacting electronic liquid, and $K < 0.5$ denotes the limit of strong (repulsive) interactions). The signature (and tunability) of strong electronic correlations in Luttinger liquids, across 2D TI material systems, not only confirms the strictly 1D nature of the helical state, but further characterize it as a potential host platform for non-Abelian parafermions [16, 17] (see section 5.1).

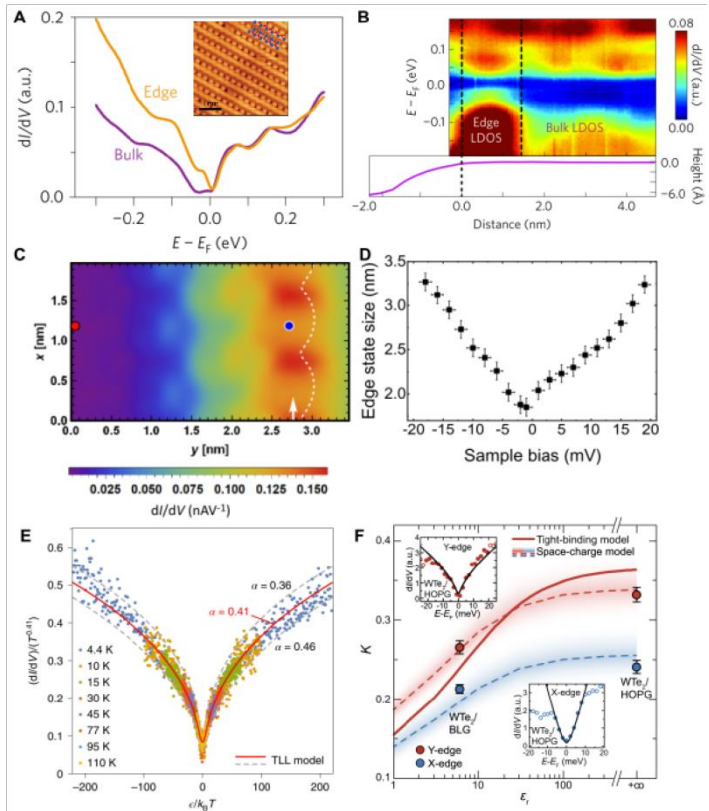

**Figure 1.** STM/STS demonstration of the QSH insulators. A) Comparison of STS spectra of the QSH edge and bulk in monolayer 1T'-WTe$_2$. B) STS line spectra across monolayer edge into the 2D bulk, showing a clear transition from conductive edge states into insulating bulk [5]. C) Conductance mapping, visualizing the edge state in bismuthene [6]. D) Spatial extent of the WTe$_2$ edge state as a function of sample bias [3]. E) Pseudogap signature, demonstrating the presence of a helical TomonagaLuttinger liquid (TLL) formed at the QSH edge in bismuthene, showing universal scaling of the LDOS to extract the Luttinger parameter [6]. F) Demonstration of the effective tuning of many-body interactions in the helical Luttinger liquid in WTe$_2$ [7].

**Current and Future Challenges**

Despite the various experimental reports of insulating bulk and metallic edge states, a persistent challenge remains to provide more direct evidence for bulk-boundary correspondence and the helical character of the edge. More so, the edge states in QSH insulators are expected to be topologically protected from scattering off non-magnetic disorder. Given that transport experiments have shown strong scattering at mean free paths of ∼100 nm (e.g., in WTe$_2$ [18]), advanced scanning probe investigations will be needed to unravel the microscopic origin of scattering processes and to verify the much-coveted topological protection.

As an indirect proof of helicity may serve the behavior of the edge in its superconducting state, either intrinsic, or induced by proximity to an s-wave superconductor. Here, an interplay of topology and strong correlations is predicted to give rise to non-trivial superconducting pairing. Similar to the physics of 1D helical electrons in proximitized Rashba nanowires [19–21], theoretical studies [17, 19] have suggested that the 1D helical edge, when proximity-coupled, may host Majorana bound states at the edge's endpoints. Different from Rashba nanowires, the helical state in QSH insulators does not have to be induced by an external magnetic field. However, creating endpoints in a helical edge poses a major challenge as the edges of an ideal QSH system should be described by periodic boundary conditions. Several theoretical proposals have suggested to create finite length systems by locally opening topologically trivial energy gaps at the Fermi energy [19] at whose boundaries Majorana bound states $\gamma_{1,2}$ can emerge (Figure 2A-B). Proposals include gapping a proximitized edge by either a ferromagnetic insulator or a local gate electrode.

If strong enough electronic interactions are present in a helical Luttinger liquid ($K < 1/4$), the presence of Umklapp scattering may behave as an impurity-induced two-particle correlated backscattering and open a gap near the Fermi energy. In this case, fractionalized Majorana fermions (parafermions) are expected [17, 22] at their boundaries (Figure 2C). Exceptionally strong interactions in this limit have been observed, for instance, in $WTe_2$ [7].

STM/STS, owing to its extremely high spatial and spectral resolution, may prove invaluable to probing the spectroscopic signatures of Majorana or parafermion bound states in real-space in which they appear as zero-energy modes similar to what has recently been shown for the helical edges of the higher-order TI bismuth [23] (Figure 2D-E).

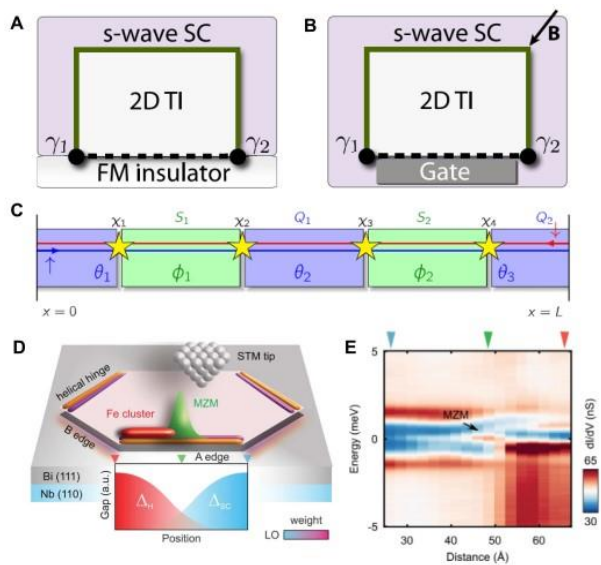

**Figure 2.** Emergence of Majorana fermions/parafermions. A, B) Proposed device configuration for trapping Majorana bound states along the helical edge state of QSH insulators [19]. The MZMs can be trapped by using A) a ferromagnetic insulator or B) gating. C) Emergence of non-Abelian parafermions hosted at the boundaries of sections along the helical edge that are either gapped by superconductivity or by strong interactions leading to Umklapp scattering [17]. D, E) Detecting Majorana zero modes (MZM) at a superconducting helical edge by STS/STS [23].

**Advances in Science and Technology to Meet Challenges**

To confirm helicity (i.e. spin-momentum locking), several theoretical studies [24–26] have proposed to employ spin-polarized STM (SP-STM) to inject spin-polarized electrons from a magnetic tunneling tip locally into a QSH edge (Figure 3A). This way, a fractionalized spin current may be generated along the edge due to the difference in the tip's and edge's spin polarization ($\theta$). This spin current can in principle be detected by electrical contacts along the edge, provided that spin-momentum locking directs the current towards one of the two contacts ($C_1$ or $C_2$). Figure 3B plots the expected timeresolved current response for two different values of $\theta$ based on current pulses from the STM tip.

The robustness of the edge state against scattering can be investigated by Fourier transform scanning tunneling spectroscopy (FT-STS) [27], which analyzes local density modulations in real space arising from quasiparticle scattering and interference (QPI). Analyzing these QPI patterns in Fourier space, equienergy surfaces (or points in 1D) can be resolved. Similar techniques have been demonstrated for spinful Luttinger liquids in 1D systems [28, 29] but not so far for helical 2D TI edge states [29] in which QPI would only be expected in the presence of back-scattering. Alternatively, charge transport may be resolved at the nanoscale using four-probe STM as recently shown [30]. As shown in Figure 3C, a four-probe resistance can be measured using four STM tips in point contacts to the QSH edge, allowing to infer the length scale over which the QSH conductance remains quantized to its ballistic limit of $2e^2/h$.

Towards the creation and trapping of Majorana fermions or parafermions, material and device developments are much needed in order to engineer the required superconducting and magnetic 2D TI heterostructures. Success in inducing superconductivity has recently been reported for 2D TI edge states [8, 9], and efforts are ongoing to generate suitable topological defects as to gap out sections of the edge and trap Majorana or parafermions.

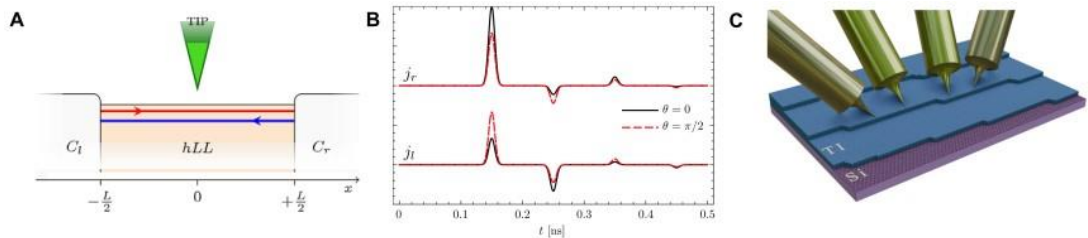

**Figure 3.** Tests of helicity and topological protection. A, B) Proposed spin-polarized STM-based detection of helicity in a TLL [25]. It shows time-resolved current response for two different fractionalized spin current pulses from the STM. C) Schematic of a four-probe STM contacting the edge of a 2D TI to confirm conductance quantization [30].

**Concluding Remarks**

Scanning tunneling microscopy has proven an invaluable tool for the investigation of 2D TIs with atomic resolution in real-space. In particular, it has shown to provide supporting evidence of bulkboundary correspondence, being able to detect the concurrence of bulk gaps and metallic edge states. More so, STM/STS have also revealed the signature of electronic correlations characterizing the edge state as a strongly interacting helical Luttinger liquid. At even lower (millikelvin) temperatures, STM/STS with its extremely high spatial and spectral resolution STM/STS promises detection of Majorana and parafermion bound states – the next frontier in 2D TI research – provided suitable material heterostructures and topological defects can be engineered.

**Acknowledgements**

This research is supported by the National Research Foundation (NRF) Singapore, under the Competitive Research Programme "Towards On-Chip Topological Quantum Devices" (NRF-CRP212018-0001) with further support from the Singapore Ministry of Education (MOE) Academic

Research Fund Tier 3 grant (MOE2018-T3-1-002) "Geometrical Quantum Materials". BW acknowledges a Singapore National Research Foundation (NRF) Fellowship (NRF-NRFF2017-11).

**References**

[1]  M. S. Lodge, S. A. Yang, S. Mukherjee, and B. Weber, "Atomically Thin Quantum Spin Hall Insulators," *Advanced Materials*, vol. 33, no. 22, p. 2008029, Jun. 2021, doi: 10.1002/adma.202008029.

[2]  J. L. Collins *et al.*, "Electric-field-tuned topological phase transition in ultrathin Na3Bi," *Nature*, vol. 564, no. 7736, pp. 390–394, Dec. 2018, doi: 10.1038/s41586-018-0788-5.

[3]  Y. Maximenko *et al.*, "Nanoscale studies of electric field effects on monolayer 1T'-WTe2," *NPJ Quantum Mater*, vol. 7, no. 1, pp. 1–6, Mar. 2022, doi: 10.1038/s41535-022-00433-x.

[4]  P. Bampoulis *et al.*, "Quantum spin Hall states and topological phase transition in germanene," *Phys Rev Lett*, 2023, doi: 10.21203/rs.3.rs-2135721/v1.

[5]  S. Tang *et al.*, "Quantum spin Hall state in monolayer 1T'-WTe2," *Nat Phys*, vol. 13, no. 7, pp. 683–687, 2017, doi: 10.1038/nphys4174.

[6]  R. Stühler *et al.*, "Tomonaga-Luttinger liquid in the edge channels of a quantum spin Hall insulator," *Nat Phys*, vol. 16, no. 1, pp. 47–51, Jan. 2020, doi: 10.1038/s41567-019-0697-z.

[7]  J. Jia *et al.*, "Tuning the many-body interactions in a helical Luttinger liquid," *Nature Communications 2022 13:1*, vol. 13, no. 1, pp. 1–7, Oct. 2022, doi: 10.1038/s41467-02233676-0.

[8]  F. Lüpke *et al.*, "Proximity-induced superconducting gap in the quantum spin Hall edge state of monolayer WTe2," *Nat Phys*, vol. 16, no. 5, pp. 526–530, 2020, doi: 10.1038/s41567-0200816-x.

[9]  W. Tao *et al.*, "Multiband superconductivity in strongly hybridized 1T'–WTe2/NbSe2 heterostructures," *Phys Rev B*, vol. 105, no. 9, p. 94512, 2022, doi: 10.1103/physrevb.105.094512.

[10] S. H. Kim, K. H. Jin, J. Park, J. S. Kim, S. H. Jhi, and H. W. Yeom, "Topological phase transition and quantum spin Hall edge states of antimony few layers," *Scientific Reports 2016 6:1*, vol. 6, no. 1, pp. 1–7, Sep. 2016, doi: 10.1038/srep33193.

[11] S.-Y. Zhu *et al.*, "Evidence of Topological Edge States in Buckled Antimonene Monolayers," *Nano Lett*, vol. 19, no. 9, pp. 6323–6329, 2019, doi: 10.1021/acs.nanolett.9b02444.

[12] C. Ghosal, M. Gruschwitz, J. Koch, S. Gemming, and C. Tegenkamp, "Proximity-Induced Gap Opening by Twisted Plumbene in Epitaxial Graphene," *Phys Rev Lett*, vol. 129, no. 11, Sep. 2022, doi: 10.1103/PhysRevLett.129.116802.

[13] F. Reis *et al.*, "Bismuthene on a SiC substrate: A candidate for a high-temperature quantum spin Hall material," *Science (1979)*, vol. 357, no. July, pp. 287–290, 2017.

[14] C. L. Kane and E. J. Mele, "Quantum spin Hall effect in graphene," *Phys Rev Lett*, vol. 95, no. 22, pp. 1–4, 2005, doi: 10.1103/PhysRevLett.95.226801.

[15] M. Bieniek, J. I. Väyrynen, G. Li, T. Neupert, and R. Thomale, "Theory of glide symmetry protected helical edge states in a WTe2 monolayer," *Phys Rev B*, vol. 107, no. 19, pp. 1–18, 2023, doi: 10.1103/PhysRevB.107.195105.

[16] F. Zhang and C. L. Kane, "Time-reversal-invariant Z4 fractional Josephson effect," *Phys Rev Lett*, vol. 113, no. 3, 2014, doi: 10.1103/PhysRevLett.113.036401.

[17] C. P. Orth, R. P. Tiwari, T. Meng, and T. L. Schmidt, "Non-Abelian parafermions in timereversal-invariant interacting helical systems," *Phys Rev B*, vol. 91, no. 8, p. 81406, 2015, doi: 10.1103/PhysRevB.91.081406.

[18] C. Liu, D. Culcer, Z. Wang, M. T. Edmonds, and M. S. Fuhrer, "Helical Edge Transport in Millimeter-Scale Thin Films of Na3Bi," *Nano Lett*, vol. 20, no. 9, pp. 6306–6312, 2020, doi: 10.1021/acs.nanolett.0c01649.

[19] J. Alicea, "New directions in the pursuit of Majorana fermions in solid state systems," *Reports on Progress in Physics*, vol. 75, no. 7. 2012. doi: 10.1088/0034-4885/75/7/076501.

[20] C. Reeg, O. Dmytruk, D. Chevallier, D. Loss, and J. Klinovaja, "Zero-energy Andreev bound states from quantum dots in proximitized Rashba nanowires," *Phys Rev B*, vol. 98, no. 24, pp. 1–12, 2018, doi: 10.1103/PhysRevB.98.245407.

[21] S. Das Sarma, "In search of Majorana," *Nat Phys*, vol. 19, no. 2, pp. 165–170, 2023, doi: 10.1038/s41567-022-01900-9.

[22] C. Wu, B. A. Bernevig, and S. C. Zhang, "Helical liquid and the edge of quantum spin hall systems," *Phys Rev Lett*, vol. 96, no. 10, pp. 1–4, 2006, doi: 10.1103/PhysRevLett.96.106401.

[23] B. Jäck, Y. Xie, J. Li, S. Jeon, B. A. Bernevig, and A. Yazdani, "Observation of a Majorana zero mode in a topologically protected edge channel," *Science (1979)*, vol. 364, no. 6447, pp. 1255–1259, 2019, doi: 10.1126/science.aax1444.

[24] S. Das and S. Rao, "Spin-Polarized Scanning-Tunneling Probe for Helical Luttinger Liquids," *Phys Rev Lett*, vol. 106, p. 236403, 2011, doi: 10.1103/PhysRevLett.106.236403.

[25] A. Calzona, M. Carrega, G. Dolcetto, and M. Sassetti, "Reprint of: Transient dynamics of spinpolarized injection in helical Luttinger liquids," *Physica E Low Dimens Syst Nanostruct*, vol. 82, pp. 229–235, Aug. 2016, doi: 10.1016/j.physe.2016.02.035.

[26] D. N. Aristov and R. A. Niyazov, "Spin-polarized tunneling into helical edge states: Asymmetry and conductances," *EPL*, vol. 117, p. 27008, 2017, doi: 10.1209/0295-5075/117/27008.

[27] L. Simon, C. Bena, F. Vonau, M. Cranney, and D. Aubel, "Fourier-transform scanning tunnelling spectroscopy: The possibility to obtain constant-energy maps and band dispersion using a local measurement," *J Phys D Appl Phys*, vol. 44, no. 46, 2011, doi: 10.1088/00223727/44/46/464010.

[28] J. Lee, S. Eggert, H. Kim, S. J. Kahng, H. Shinohara, and Y. Kuk, "Real space imaging of onedimensional standing waves: direct evidence for a luttinger liquid," *Phys Rev Lett*, vol. 93, no. 16, 2004, doi: 10.1103/PhysRevLett.93.166403.

[29] T. Zhu *et al.*, "Imaging gate-tunable Tomonaga–Luttinger liquids in 1H-MoSe2 mirror twin boundaries," *Nat Mater*, vol. 21, no. 7, pp. 748–753, Jun. 2022, doi: 10.1038/s41563-02201277-3.

[30]    A. Leis *et al.*, "Probing Edge State Conductance in Ultra-Thin Topological Insulator Films," *Adv Quantum Technol*, vol. 5, no. 9, Sep. 2022, doi: 10.1002/qute.202200043.

## 5.1 Interacting quantum channel engineering from quantum spin Hall edge modes

Maciej Bieniek[1,2], Jukka I. Väyrynen[3], Ronny Thomale[1]

(1) Institut für Theoretische Physik und Astrophysik, Universität Würzburg, 97074 Würzburg, Germany
(2) Department of Theoretical Physics, Wrocław University of Science and Technology, Wybrzeże Wyspiańskiego 27, 50-370 Wrocław, Poland
(3) Department of Physics and Astronomy, Purdue University, West Lafayette, Indiana 47907 USA

**Status**

Strongly interacting electrons confined to one dimension (1D) form a Tomonaga-Luttinger liquid (TLL), whose properties are distinct from higher-dimensional Fermi liquids. A long-standing problem is how to achieve such truly 1D carrier confinement. One avenue is to fabricate nanowires, nanotubes, or atomic wires in which energy levels are quantized in two out of three dimensions. Another method is to use a strong magnetic field to push electrons to the edge of two dimensional (2D) system, creating chiral TLL in samples in integer or fractional quantum Hall regimes. An even simpler method is to utilize natural 2D materials with edge states forming without magnetic field. Ultimately, one can leverage the non-trivial topology which allows edge states to be protected by invariants, as in Chern or quantum spin Hall insulators [1]. Nowadays, thanks to many advances in 2D atomically thin topological materials, the atlas of systems exhibiting interesting edge phenomena has grown immensely, with notable examples of bismuthene [2] and WTe2 [3–5]. These systems allow for the study of the 1D Fermi sea via scanning tunneling microscopy (STM) techniques, greatly enhancing possibility of experimental probing of TLL properties beyond transport and angle- resolved photoemission spectroscopy (ARPES).

One of the most promising aspects of the interplay of topology and strong correlations is the exotic physics that arises when these materials are taken to the superconducting state, either by proximity effect or by intrinsic pairing. The strongly correlated counterpart of the topological insulator (TI) edge is an engineered helical fractional quantum Hall edge. Proximitizing such an edge with an s-wave superconductor should allow one to capture a "fractionalized Majorana" known as parafermion [6]. This non-Abelian anyon can be used to perform topologically protected Clifford gates. A network of parafermions can also be used to construct emergent Fibonacci anyons which enable a topologically protected universal quantum computation.

**Current and Future Challenges**

We begin with a discussion of experimental challenges. The first one is the identification of suitable material platforms that will allow the study of 1D channels using different experimental techniques such as STM or transport measurements. STM techniques are well suited to study atomically thin materials (such as monolayer WTe2) where the TLL of interest is accessible with a surface probe. An important challenge is the experimental identification of the role of dielectric environment on screening of electron-electron interactions in various TLL systems. Focusing on challenges in STM studies, it is necessary to obtain position-independent tunneling spectra, which are now usually obtained from highly disordered edges. There, it would be interesting to clarify relation between finite-size quantization and TLL physics [7].

Molecular beam epitaxy grown monolayers allow STM studies in sufficiently clean in regions of the sample. However, the small size of clean areas makes transport experiments challenging. For this reason, much bigger flakes have to be used, which are highly inhomogeneous at the scale of the system due to the exfoliation process. In exfoliated samples disorder in the bulk is currently a challenge, as seen in the case of WTe2. The approximate conductance quantization was only observed there in short edges of order 100 nm [8, 9]. This can be contrasted with HgTe semiconductor quantum well where gate-training can increase the average conductance near ideal value even in samples longer than 10μm [10]. We note that disorder may mask correlations in transport, but perhaps less so in STM studies. Different crystallographic orientations or atomic edge terminations can also affect TLL properties, as described in our recent work [4].

Moving to theoretical challenges, it remains formidable to simulate large, strongly interacting systems. Starting from ab initio simulations, it is already difficult to build reliable mean-field level models of realistic samples, which may have complicated electronic structure, especially at material boundaries. An additional complication is the description of mean-field ground states that are not simple band insulators, but rather charge density wave or excitonic insulators. These theoretical models need to be then sufficiently simplified to serve as a basis for interacting theories, such as extended Hubbard model-type models. Solving such many-body problems for realistic systems, such as quantum spin Hall channels, is another challenge. Exact diagonalization, quantum Monte Carlo, and renormalization group methods are usually limited to a few particles and a restricted number of possible excitations. It is also challenging to construct quasi-1D systems, in which quantum spin Hall edge states with realistic wave functions can be simulated. Those can serve further as a building blocks for higher dimensional integer and fractional topological quantum states [11].

**Advances in Science and Technology to Meet Challenges**

From an experimental perspective, one of the important advances necessary for the field to progress is the identi- fication and growth of systems that achieve 1D confinement with reduced disorder both in the bulk and on the edge where the TLL lives. In the case of monolayer topological systems, it is critical to obtain atomically well-defined edges, similar to the progress already achieved in the synthesis of graphene with well-defined boundaries and extremely low disorder. It is also an outstanding challenge to obtain clean enough samples to have ballistic helical edge conduction over long distances (of the order of 1 μm). The next necessary step is to identify proper substrates that allow for the study of the role of the dielectric environment on the strength of electron-electron interactions. To address Majorana/ parafermion physics it is pivotal to find appropriate superconducting substrates [12]. On another front, the synthesis of larger samples would allow performing combined transport, STM, and ARPES studies, providing access to a variety of TLL properties. Additionally, the gate tunability of the Fermi level of such systems would allow for the probing of potential nonlinearities.

From a theoretical perspective, there are several necessary advances required to build material-realistic theories of Luttinger liquids. The first is the identification of minimal models that accurately describe the dispersion of 1D bands, along with proper scales where 'high energy' begins to set in. In realistic 2D platforms this means an appropriate description of band edges. An even more challenging forefront is the developement of theoretical and numerical methods that would enable the study of strongly correlated quasi-2D systems. Some potential avenues for progress involve machine learning to accelerate convergence of standard techniques, such as exact diagonalization, quantum Monte Carlo, density matrix renormalization group or functional renormalization group. Combining these with the development of algorithms that could use 'classical' exa-scale supercomputers may offer a window into the realistic modelling of TLL properties. Another exciting front is the development of algorithms that would utilize the already available noisy intermediate-scale quantum computers.

**Concluding Remarks**

To conclude, the study of the 1D electronic phenomena remains a fascinating field of research, both for fundamental understanding and for potential applications such as highly anticipated Majorana-based fault tolerant quantum computing. With the number of 2D topological materials slowly but steadily increasing, the combination of low dimensionality, topology, and strong interactions, without the necessity of high magnetic fields and ultra-low temperatures, is expected to usher a new era in the study of 1D quantum liquids.

**Acknowledgements**

*We acknowledge support from the Deutsche Forschungsgemeinschaft (DFG, German Research Foundation) through QUAST FOR 5249-449872909 (Project P3). The work in Würzburg is further supported by the Deutsche Forschungsgemeinschaft (DFG, German Research Foundation) through Project-ID 258499086-SFB 1170 and the Würzburg- Dresden Cluster of Excellence on Complexity and Topology in Quantum Matter – ct.qmat Project-ID 390858490-EXC 2147. J.I.V. acknowledges support by the US Department of Energy (DOE) Office of Science through the Quantum Science Center (QSC, a National Quantum Information Science Research Center.*

**References**

[1] J. Strunz, J. Wiedenmann, C. Fleckenstein, L. Lunczer, W. Beugeling, V. L. Müller, P. Shekhar, N. T. Ziani, S. Shamim, J. Kleinlein, H. Buhmann, B. Trauzettel, and L. W. Molenkamp, "Interacting topological edge channels", *Nature Physics* 16, 83 (2020).

[2] R. Stühler, F. Reis, T. Müller, T. Helbig, T. Schwemmer, R. Thomale, J. Schäfer, and R. Claessen, "Tomonaga–Luttinger liquid in the edge channels of a quantum spin Hall insulator", *Nature Physics* 16, 47 (2020).

[3] J. Jia, E. Marcellina, A. Das, M. S. Lodge, B. Wang, D.-Q. Ho, R. Biswas, T. A. Pham, W. Tao, C.-Y. Huang, H. Lin, A. Bansil, S. Mukherjee, and B. Weber, "Tuning the many-body interactions in a helical Luttinger liquid", *Nature Communications* 13, 6046 (2022).

[4] M. Bieniek, J. I. Väyrynen, G. Li, T. Neupert, and R. Thomale, "Theory of glide symmetry protected helical edge states in a WTe2 monolayer", *Physical Review B 107*, 195105 (2023).

[5] Y.-Q. Wang, M. Papaj, and J. E. Moore, Breakdown of helical edge state topologically protected conductance in time- reversal-breaking excitonic insulators (2023), arXiv:2305.09202.

[6] J. Alicea and P. Fendley, "Topological Phases with Parafermions: Theory and Blueprints", *Annual Review of Condensed Matter Physics* 7, 119 (2016).

[7] T. Zhu, W. Ruan, Y.-Q. Wang, H.-Z. Tsai, S. Wang, C. Zhang, T. Wang, F. Liou, K. Watanabe, T. Taniguchi, J. B. Neaton, A. Weber-Bargioni, A. Zettl, Z. Q. Qiu, G. Zhang, F. Wang, J. E. Moore, and M. F. Crommie, "Imaging gate-tunable Tomonaga–Luttinger liquids in 1H-MoSe2 mirror twin boundaries", *Nature Materials* 21, 748 (2022).

[8] Z. Fei, T. Palomaki, S. Wu, W. Zhao, X. Cai, B. Sun, P. Nguyen, J. Finney, X. Xu, and D. H. Cobden, "Edge conduction in monolayer WTe2", *Nature Physics* 13, 677 (2017).

[9] S. Wu, V. Fatemi, Q. D. Gibson, K. Watanabe, T. Taniguchi, R. J. Cava, and P. Jarillo-Herrero, "Observation of the quantum spin Hall effect up to 100 kelvin in a monolayer crystal", *Science* 359, 76 (2018).

[10] L. Lunczer, P. Leubner, M. Endres, V. L. Müller, C. Brüne, H. Buhmann, and L. W. Molenkamp, "Approaching Quantization in Macroscopic Quantum Spin Hall Devices through Gate Training", *Physical Review Letters* 123, 047701 (2019).

[11] T. Neupert, C. Chamon, C. Mudry, and R. Thomale, "Wire deconstructionism of two-dimensional topological phases", *Physical Review B* 90, 205101 (2014).

[12] W. Tao, Z. J. Tong, A. Das, D.-Q. Ho, Y. Sato, M. Haze, J. Jia, Y. Que, F. Bussolotti, K. E. J. Goh, B. Wang, H. Lin, A. Bansil, S. Mukherjee, Y. Hasegawa, and B. Weber, "Multiband superconductivity in strongly hybridized 1T'–WTe2/NbSe2 heterostructures", *Physical Review B* 105, 094512 (2022).

## 5.2 The Topological Quantum Field-Effect Transistor for Low-Voltage Switching

Michael S. Fuhrer[1,2], Dimitrie Culcer[3,4], Bhaskaran Muralidharan[5], Muhammad Nadeem[6,7]

[1]School of Physics and Astronomy, Monash University, Clayton, Victoria 3800, Australia
[2]ARC Centre of Excellence in Future Low-Energy Electronics Technologies (FLEET), Monash University, Clayton, Victoria 3800, Australia
[3]School of Physics, University of New South Wales, Sydney 2052, Australia
[4]ARC Centre of Excellence in Future Low-Energy Electronics Technologies (FLEET), University of New South Wales, Sydney 2052, Australia
[5]Department of Electrical Engineering, Indian Institute of Technology Bombay, Powai, Mumbai 400076, India
[6]Institute for Superconducting and Electronic Materials (ISEM), Australian Institute for Innovative Materials (AIIM), University of Wollongong, Wollongong, New South Wales 2525, Australia
[7]ARC Centre of Excellence in Future Low-Energy Electronics Technologies (FLEET), University of Wollongong, Wollongong, New South Wales 2525, Australia

**Status**

Growth in demand for computation is significantly outpacing the efficiency gains provided by Moore's Law. As a result, the energy used in computation is doubling approximately every three years, and will be strongly limited by global energy production in the next two decades unless new solutions are found to compute at lower energy [1]. The operating voltage, and hence switching energy, of a conventional MOSFET is limited by thermal activation of charge carriers over the barrier in the channel $\leq e\Delta V_g$ created by a gate voltage $\Delta V_g$, resulting in a subthreshold swing $S \geq (kT/e)\ln10 \approx 60$ mV/decade at room temperature, a limit termed "Boltzmann's tyranny". Strategies to beat Boltzmann's tyranny to switch at lower voltage are an active area of research.

Figure 1 shows the topological quantum field-effect transistor (TQFET) schematically. The TQFET uses an electric field to alter the bandgap of a 2D topological insulator (quantum spin Hall) material near the quantum critical point between topological and trivial insulator. An "off" state is generated by the opening of the trivial bandgap. Strong spin-orbit coupling due to the Rashba interaction $\lambda_R$ (inherent to topological insulators) enhances the change in bandgap with electric field (see Fig. 1d), such that a barrier $\geq e\Delta V_g$ can be created by application of a potential difference $\Delta V_g$ across the channel, resulting in a subthreshold $S \leq (kT/e)\ln10$ [2].

When a negative capacitor (a ferroelectric operated in a non-hysteretic regime) is placed in series with a positive capacitance (dielectric, in this case the channel material of the TQFET), the electric field is amplified in the dielectric. This effect has been proposed to improve the performance of conventional MOSFETs[3], however in an ideal MOSFET the change in electric field in the subthreshold regime is *zero* (the channel potential follows the gate potential) hence the negative capacitance does not lower the subthreshold swing [4]. Because the TQFET is inherently driven by an electric field across the bulk which is always insulating, it can take full advantage of *negative capacitance* in the gate stack[5]. The operation of the negative-capacitance TQFET (NC-TQFET) is shown schematically in Fig. 2.

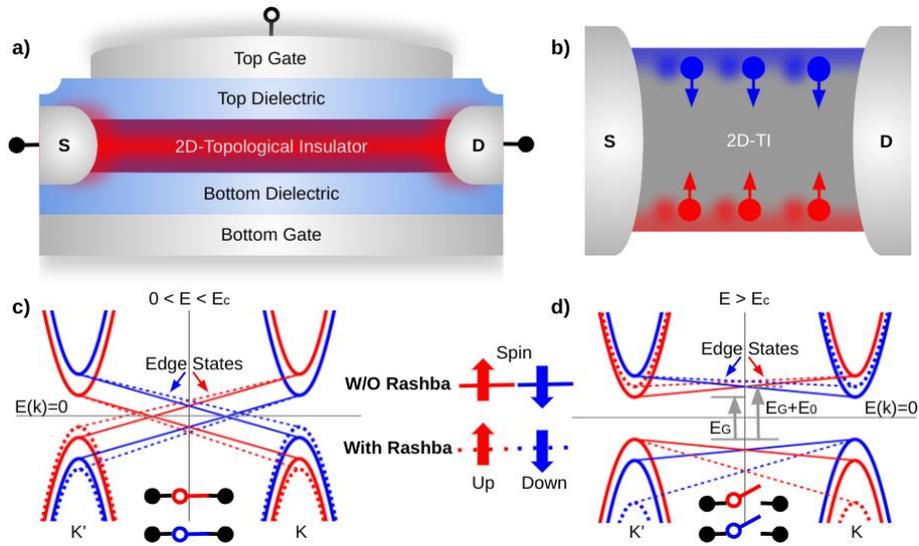

Figure 1. Topological quantum field effect transistor. (a) Structure of the device: a 2D topological insulator is sandwiched between top and bottom dielectrics and gate electrodes, and contacted by source (S) and drain (D) electrodes. (b) Schematic of a topological insulator channel with edge state conduction in the "on" state. (c,d) Band diagrams in "on" (topological insulator, c) and "off" (conventional insulator, d) states. When the gate electric field is less than a critical value, $0 < E < E_c$, topological insulator hosts gapless helical conducting edge channels with a minimum value of the quantized conductance 2 $e^2/h$. When the gate electric field exceed a critical value, $E > E_c$, a topological insulator enters into a normal insulating regime in which the conductance is exponentially suppressed by the energy barrier $E_G$. Such electric field switching is accompanied by the topological quantum field effect via the Rashba effect, which enhances the topological phase transition driven by a gate electric field, increasing the barrier to $E_G + E_0$ and reducing the subthreshold swing. Here solid and dashed "spin-up" (red) and "spin-down" (blue) states represent band dispersions without and with Rashba spin-orbit interaction respectively, while $E_0$ represents a shift in nontrivial (c) and trivial (d) band gap $E_G$ due to the topological quantum field effect.

**Current and Future Challenges**

Significant challenges remain in understanding the electronic structure and band translation physics of topological insulators in electric fields. Near the zero of energy, the edge state conductance is quantized, and this quantization is protected by the topology of the bulk band structure. The top gate electric field induces a topological phase transition in which the conductance drops from $\frac{2e^2}{h}$ (where $e$ is the elemental charge, $h$ is Planck's constant) to zero as the system passes from the quantum spin Hall (QSH; "on") into the trivial regime ("off"). The topological phase transition is captured by a simple next-nearest-neighbour tight-binding model, in which the non-trivial band gap is opened by the intrinsic spin-orbit coupling, and the top gate induces an energy difference between sub-lattices. These simple considerations yield *S* = (*kT*/*e*)ln10 since the nontrivial (trivial) band gap decreases (increases) linearly with the electric field.

To capture the full slope as a function of the top gate field one must incorporate the Rashba spin-orbit interaction that results from mirror symmetry breaking, represented by nearest neighbour hopping. At the microscopic level the Rashba interaction arises from the mixing of σ and π bands brought about by the intrinsic atomic spin-orbit coupling and the Stark effect. The Rashba interaction reduces the nontrivial band gap opened by the intrinsic spin-orbit coupling and enhances the trivial band gap opened by the top gate. In the process of electric field switching, the strength of the Rashba interaction also varies linearly with the top gate field, resulting in an overall nonlinear dependence of the gap on the top gate field, and yielding *S* ≤ (*kT*/*e*)ln10. It is this change in the switching properties that we denote by the topological quantum field effect. The effect provides tunable parameters for controlling the sub-threshold swing in a TQFET rather than relying purely on the gate capacitance mechanism.

The precise value of S will be strongly material-dependent. A small sub-threshold swing requires materials whose gap increases at a fast rate as a function of the top gate voltage, ruling out bilayer graphene, whose gap increases more slowly than initial theoretical predictions. For eligible materials a challenge will be to determine the exact value of S using a model that includes screening, strain and the detailed gate architecture.

While band translation physics indicates the possibility a steep sub-threshold operation near the quantum critical point, a detailed device simulation sheds more light on the intricacies of the actual device operation. Quantum transport may be analysed for a three terminal device structure comprising the source-drain electrodes and the inclusion of the gate electrostatics. The terminal currents $I_D$, are evaluated at an applied bias, $V_{DS}$, across the source and drain, at a given gate voltage, $V_G$. From this, one can ultimately calulate the sub-threshold swing via the $log(I_D) - V_G$ characteristics at the operating bias. Sufficient advances in device modelling have certainly permitted an accurate and predictive device simulation platform for Si-based nano-transistor structures [6].

In this context, quantum transport modelling using the non-equilibrium Green's function (NEGF) platform, has in the past two decades provided a strong underpinning for nanoscale device modelling, especially in the context of Si-based nanoscale transistors. Details of computational modelling aspects of 2D-topological electronics are spelt out in section 5.3.

A recent detailed quantum transport simulation has adapted the NEGF formalism using a parametrized tight binding framework [7] on a 2D-Xene nanoribbon as a model channel with Rashba interactions, coupled with dual gate electrostatics. It has indeed computationally demonstrated a viable TFET device operation around the quantum critical point, with the application of gate fields. The nano-ribbon geometry along with its edge-termination appropriately accounts for the topological edge states in the topological phases. The sub-threshold characteristics modelled in detail the topological phase transition that can be induced via gating.

Further advancements in computational modelling must encompass various developments along the following lines: a) detailed electronic structure via the Wannierized tight binding formalism adapted to the nanoribbon structure [8] that takes care of the electronic structure in detail, which should be coupled with the NEGF formalism in this basis, b) Detailed electrostatics along the lines of a detailed self-consistent simulation platform [6], c) inclusion of defects and dephasing. Once such a platform is advanced, one can then add the negative capacitance effect via the inclusion of the ferroelectric oxide physics, as explored in related works in the case of a Si-channel NC-MOSFET [3]. We envisage that aforementioned device modelling advancements in parallel will truly complement and accelerate experimental developments.

**Advances in Science and Technology to Meet Challenges**

Significant experimental challenges still exist to realisation of topological transistors. Experimental realisations of electric field switching of the topological state have so far been limited to $Na_3Bi$ [9] and germanene [10], in a geometry utilizing a scanning tunnelling microscope probe to simultaneously apply a gate electric field and measure the band gap via tunnelling spectroscopy. Electric-field switching of a topological transport current (i.e. a topological transistor) has not yet been demonstrated. Identifying suitable topological materials which can be integrated with gate dielectrics and electrodes is necessary to advance the study of topological transistors.

Realisation of the NC-TQFET concept will require integration of topological insulators with ferroelectric layers (see Fig. 2). Top-down fabrication strategies would require identifying lattice-matched topological and ferroelectric materials which could be grown sequentially, which will severely (and perhaps prohibitively) limit materials choices. The recent advent of van der Waals heterostructures could offer a solution, enabling integration of widely disparate materials. There are already experimentally realised examples of van der Waals topological insulators (e.g. 1T'-WTe$_2$) and ferroelectrics (e.g. α-In$_2$Se$_3$), as well as emergent ferroelectricity in vdW heterostructures of non-ferroelectric constituents. Another possibility is to use free-standing oxide membranes prepared epitaxially on sacrificial substrates such as Sr$_3$Al$_2$O$_6$.

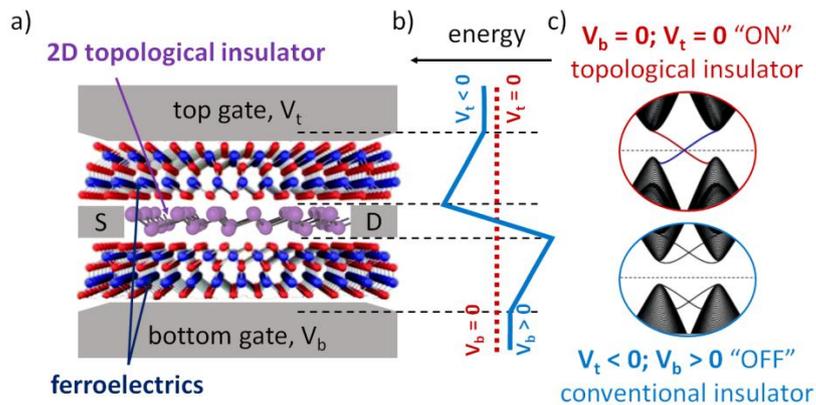

Figure 2. Schematic of negative capacitance topological quantum field-effect transistor. (a) Structure of the device: a 2D topological insulator is sandwiched between two ferroelectrics and top and bottom gate electrodes with voltages $V_t$ and $V_b$ respectively, and contacted by source (S) and drain (D) electrodes. (b) Electrostatic potential energy as a function of distance across the device in "on" (red) and "off" (blue) states. (c) Band diagrams in "on" (topological insulator, red) and "off" (conventional insulator, blue) states. Adapted from [5].

**Concluding Remarks**

The growing energy use in computing technologies has created an urgent need for transistor concepts capable of switching at lower voltage than conventional silicon CMOS. Recently the prospect of low-voltage switching based on topological insulators has emerged as a potentially transformative application of topological materials. Increased theoretical understanding of electric field effects on the band structure of topological materials, as well as more realistic device modelling are needed to refine the topological transistor concept and identify candidate materials. Integration of topological materials with dielectrics, ferroelectrics, and electrodes is required to create and benchmark novel transistor structures; van de Waals heterostructures offer a particularly promising route to the integration of these disparate materials.

**Acknowledgements**

MSF, DC, and MN are supported in part by the ARC Centre of Excellence in Future Low-Energy Electronics Technologies (CE170100039).

**References**

[1] "Decadal Plan for Semiconductors", Semiconductor Research Corporation, https://www.src.org/about/decadal-plan/.
[2] M. Nadeem, I. D. Bernardo, X. Wang, M. S. Fuhrer, and D. Culcer, *Nano Letters* **21**, 3155 (2021).
[3] S. Salahuddin & S. Datta, *Nano Letters* **8**, 405 (2008).
[4] W. Cao and K. Banerjee, *Nature Communications* **11**, 196 (2020).
[5] M.S. Fuhrer, M.T. Edmonds, D. Culcer, M. Nadeem, X. Wang, N.V. Medhekar, Y. Yin, J.H. Cole, "Proposal for a Negative Capacitance Topological Quantum Field-Effect Transistor," Technical Digest - International Electron Devices Meeting, IEDM Volume 2021-December, Pages 38.2.1 - 38.2.4. 2021 IEEE International Electron Devices Meeting (IEDM 2021), San Francisco 11-16 December 2021.
[6] G. Klimeck and T. Boykin, *Springer Handbook of Semiconductor Devices*, 1601-1640, (2022).

[7] S. Banerjee, K. Jana, A. Basak, M. S. Fuhrer, D. Culcer and B. Muralidharan, *Phys. Rev. Applied,* **18**, 054088, (2022).

[8] A. Lau, R. Ray, D. Varjas and A. Akhemrov, *Phys. Rev. Materials*, **3**, 054206, (2019).

[9] Pantelis Bampoulis, Carolien Castenmiller, Dennis J. Klaassen, Jelle van Mil, Yichen Liu, Cheng-Cheng Liu, Yugui Yao, Motohiko Ezawa, Alexander N. Rudenko, and Harold J. W. Zandvliet, *Physical Review Letters* **130**, 196401 (2023).

[10] James L. Collins, Anton Tadich, Weikang Wu, Lidia C. Gomes, Joao N. B. Rodrigues, Chang Liu, Jack Hellerstedt, Hyejin Ryu, Shujie Tang, Sung-Kwan Mo, Shaffique Adam, Shengyuan A. Yang, Michael S. Fuhrer & Mark T. Edmonds, *Nature* **564**, 390 (2018).

## 5.3 2D-Topological electronics – A computational perspective

Bhaskaran Muralidharan[1]
[1]Department of Electrical Engineering, Indian Institute of Technology Bombay, Powai, Mumbai 400076, India

**Status**

Topological robustness can potentially be harnessed for modern device paradigms- a field that is termed as topological electronics [Gilbert2021]. Two aspects that stem from quantum topology can be potentially tapped for emerging device paradigms: 1) the on-demand manifestation of topologically robust states via electrical means and 2) the harnessing of topological phase transitions. To understand inner device workings, provide an underpinning to experimental efforts, and ultimately predict new paradigms of the like, one needs to expand onto a predictive computational topological electronics device simulation platform.

In the minimal form, the topological device consists of the 2D-material nanoribbon subject to open boundary conditions with the contacts, typically labelled as the source and the drain, and the electrostatic environment provided by the gates. The nanoribbon geometry is crucial to visualize the bulk-boundary correspondence in real space and the functional topography of the actual device.

In order to stress the importance of computational modelling of topological devices and the conceptualization of device structures, we refer to a specific example provided in Figure 1. The central panel shows a typical 2D-Xene material lattice, a prominent feature being the buckling along the vertical axis, which provides the opportunity to open the gap via the application of a gate voltage. We also show the possible topological phases that this material can exhibit. It is now possible to harness these phases via appropriately designed device structures. Here we depict two possibilities in the left and right panels. Gating strategies (left panel) can give rise to producing symmetry states "on demand" [Jana2022]. On the other hand, the right panel shows how one can design a topological transistor, which relies on the topological phase transition, to optimize a sub-threshold operation [Nadeem2021,Banerjee2022].

In this context, quantum transport modelling using the non-equilibrium Green's function (NEGF) platform [Camsari2022], has, in the past two decades provided a strong underpinning for nanoscale device modelling, especially in the context of Si-based nanoscale transistors [Klimeck2022]. In the ballistic limit, the formalism gives an atomistic prescription to evaluate the transmission, and connects the current operator with the Landauer-Büttiker approach. Alternatively, in this limit, one can also evaluate the transmission using the Green's function via the S-matrix approach [Groth2014].

We briefly outline the pathways and challenges ahead in terms of developing a predictive and realistic device simulation platform for 2D-topological electronics.

**Current and Future Challenges**
Much of the advances in quantum transport have been in the coherent ballistic limit [Groth2014], where detailed electronic structure calculations could supplement model Hamiltonian calculations, within a reasonable computational upgrade [Lau2019]. In the current context, specifically, the issue of edge disorder and termination of 2D-nanoribbons becomes so specific that minute variations in the simulation setups [Lau2019] can give rise to immense differences in the evaluated result, making device level predictions very difficult. Furthermore, given the computational complexity of transport simulations, there are several levels of electronic structure theory that must be considered.

To test basic device paradigms, for instance, one typically starts with simpler model Hamiltonians using the single orbital tight binding frameworks, which provides an overall working principle [Jana2022,Banerjee2022].

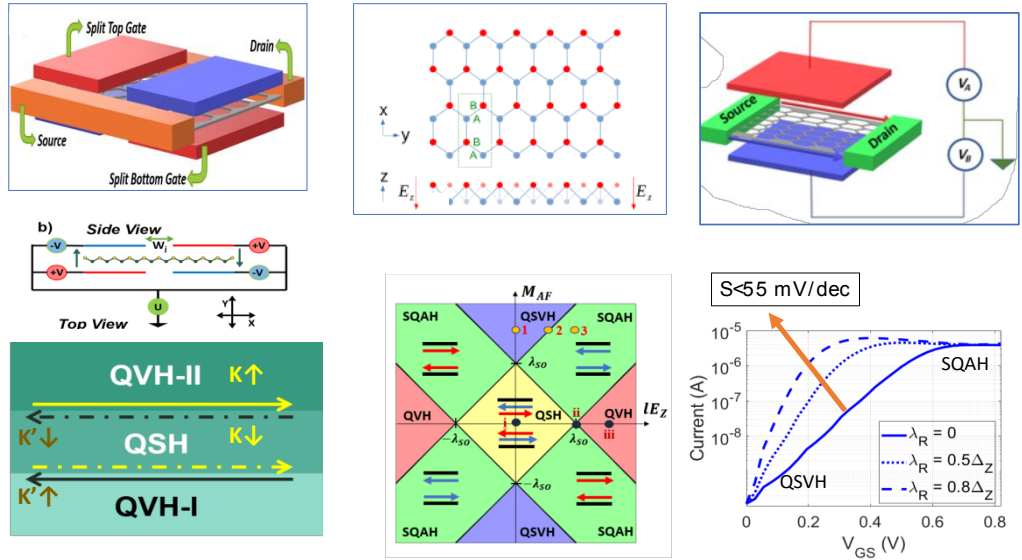

**Figure 1.** An example of a topological electronic device: (center top) typical lattice structure of 2D-Xenes. A crucial aspect here is the buckled structure that can create a gapped spectrum under the application of a vertical electric field. (center bottom) Topological phase space of such a structure. (Left) Device structure to create spin-valley locking, which effectively locks spins to valleys via an opposite polarity dual gating. The spin-valley polarized states are created via a bulk-boundary correspondence of a quantum spin Hall phase flanked between two quantum valley Hall phases. [Jana2022] (Right) Topological transistor structure which uses a dual gate to create a substantially steep subthreshold swing [Nadeem2021,Banerjee2022].

However, going beyond the 2D-Xene platform requires serious consideration of orbital chemistry and typically one has to adapt the quantum transport formulation to the Wannierized tight binding framework, where electronic structure calculations are quite well established [Lau2019,Nadeem2021].

In order to further developments in quantum transport and move toward a useful platform, one needs to move toward realistic and predictive models, which can capture detailed device intricacies including the electrostatic environment, disorder and finite temperature scattering effects [Luisier2009]. Disorder models, in general, can get very computationally intense, but can be shown to be equivalent to momentum relaxing dephasing processes [Camsari2022]. In such cases, the NEGF formalism can be advanced to capture models of dephasing using the self-consistent Born approach, which goes well beyond the ballistic limit, but can get computationally intense. It is truly a challenge to advance quantum transport with the inclusion of the full band-structure details coupled with disorder and dephasing models.

The above typically applies to dc-transport. However, given that device simulations for quantum technologies will strongly rely on hybrid quantum systems, one has to move toward effective modelling of incorporating superconductivity, the inclusion of an electromagnetic environment and circuit quantum electrodynamics. This will open several possibilities in terms of advancing topological electronic device modelling for quantum technology applications, specifically opening the door toward the atom-to-circuit device design.

**Advances in Science and Technology to Meet Challenges**
In order to sketch a roadmap for advancement, let us consider the example of a topological transistor depicted in Fig. 1. This is a case of dc-quantum transport modelling, for which a holistic platform must include the following developments. First, the development of suitable electronic structure methods for transport across various 2D-material systems, which is well integrable with NEGF, is a precursor for an effective understanding of 2D topological insulator based devices. Recognizing the importance of adding many functionalities, including high-bias transport, detailed modelling of the contacts, scattering effects and coupling to the electrostatics, such methods must be able to be encapsulated within the tight-binding real space framework. Following this, detailed numerical techniques to improve the calculation speeds of NEGF with the inclusion of scattering and dephasing effects must be developed. Having a strong platform for dc-NEGF transport will then be completed with the inclusion of the electrostatic environment, for instance the gating and complex oxide field effects that might be necessary in the case of the topological field effect transistor.

Furthermore, to model high speed operation and possibly switching transients, one has to develop the full fledged ac-transport model. Given that the dc-NEGF formalism is only a special case of the most generalized time-dependent transport equations [Camsari2022], it is a huge computational exercise to develop. A first step would be to develop ac- transport with the full force of quantum transport coupled with Maxwell's equations [Philip2018], With this, one can also analyse the high frequency operation of device structures like topological transistors. Most importantly, given the current interest in hybrid quantum devices, the development of a realistic device framework that captures the coupling to the electromagnetic environment seems to be one of a timely need.

**Concluding Remarks**
In order to construct meaningful device concepts based on what topological insulators can potentially offer, one needs to conecptualize device designs. Quantum transport based device modelling provides a strong backbone to both provide theoretical underpinning to experiments as well as provide for new device designs. However, lots of advancements on the computational framework need to be developed which include computationally efficient electronic structure extraction, efficient matrix calculations, GPU based computing and eventually development of composite solvers. All this will lead to a strong synergy between computation, experiment and theory geared toward the advancement of topological electronics for both the Beyond Moore era and the quantum technology era.

**Acknowledgements**
We gratefully acknowledge the co-authors of the papers which were discussed. We acknowledge the Visvesvaraya Ph.D. Scheme of the Ministry of Electronics and Information Technology (MEITY), Government of India, the Science and Engineering Research Board (SERB), Government of India, Grant No. CRG/2021/003102, the Ministry of Education (MoE), Government of India, Grant No. STARS/APR2019/NS/226/FS under the STARS scheme.

**References**
[Gilbert2021] M. J. Gilbert, *Comms. Phys.,* **4**, 70, (2021).
[Jana2022] K. Jana and B. Muralidharan, npj 2D Materials and Applications, **6**, 19, (2022).
[Nadeem2021] M. Nadeem, I. D. Bernardo, X. Wang, M. S. Fuhrer, and D. Culcer, *Nano Letters* **21**, 3155 (2021).
[Banerjee2022] S. Banerjee, K. Jana, A. Basak, M. S. Fuhrer, D. Culcer and B. Muralidharan, *Phys. Rev. Applied,* **18**, 054088, (2022).
[Camsari2022] K. Camsari, S. Chowdhury and S. Datta, *Springer Handbook of Semiconductor Devices*, 1583-1599, (2022).

[Klimeck2022] G. Klimeck and T. Boykin, *Springer Handbook of Semiconductor Devices*, 1601-1640, (2022).
[Groth2014] C. W. Groth, M. Wimmer, A. R. Akhmerov and X. Waintal, New. J. Phys., **16,** 063065, (2014).
[Lau2019] A. Lau, R. Ray, D. Varjas and A. Akhemrov, *Phys. Rev. Materials*, **3**, 054206, (2019).
[Luisier2009] M. Luisier and G. Klimeck, Phys. Rev. B, **80**, 155430, (2009).
[Philip2018] T. M. Philip and M. J. Gilbert, J. Comp. Elec., **17**, 934, (2018).